\documentclass[aps,prb,twocolumn,showpacs,floatfix]{revtex4-1}
\usepackage{graphicx,amsfonts,amssymb,amsmath,hyperref}

\newif\ifhyper
\hypertrue
\ifhyper
\hypersetup{
  citecolor = {green},
  colorlinks = {true}, % false
  urlcolor = {blue} % magenta
}

\usepackage{environ}
\NewEnviron{eqsplit}{%
\begin{equation}\begin{split}
  \BODY
\end{split}\end{equation}
}

\begin{document}

%AUTHOR'S MACRO
\def\Tkt{T_{\rm BKT}}

%%%%%%%%%%%%%%%%%%%%%%%%%%%%%%%%%%%%%%%%%%%

\graphicspath{{./figures/}}

\def\rhoeq{\hat\rho_{\rm eq}}

\newcommand{\marge}[1]{\marginpar{\scriptsize #1}}
\newcommand{\remarque}[1]{\marginpar{\scriptsize Remarque}{\it [#1]}}
\newcommand{\red}[1]{\textcolor{red}{#1}}
\newcommand{\blue}[1]{\textcolor{blue}{#1}}

\def\beq{\begin{equation}}
\def\eeq{\end{equation}}
\def\bleq{\begin{eqnarray}}
\def\eleq{\end{eqnarray}} 
\def\bfig{\begin{figure}}
\def\efig{\end{figure}}
\def\bline{\begin{multline}}
\def\eline{\end{multline}}
\def\bremark{\begin{quotation} \noindent \small }
\def\eremark{\end{quotation}}
\def\llbrace{\left\lbrace}
\def\rrbrace{\right\rbrace}
\def\rrangle{\right\rangle}
\def\llangle{\left\langle}
\def\lbraket{\left[}
\def\rbraket{\right]}

\newcommand{\Tr}{{\rm Tr}} 
\newcommand{\tr}{{\rm tr}} 
\newcommand{\sgn}{{\rm sgn}} 
\newcommand{\mean}[1]{\langle #1 \rangle}
\newcommand{\commu}[2]{[#1,#2]} 
\newcommand{\bra}[1]{\langle#1|}
\newcommand{\ket}[1]{|#1\rangle}
\newcommand{\braket}[2]{\langle #1|#2\rangle}
\newcommand{\dbraket}[3]{\langle #1|#2|#3\rangle}
\newcommand{\tens}[1]{\overleftrightarrow{#1}}  
\newcommand{\vac}{|{\rm vac}\rangle} 
\def\bravac{\langle{\rm vac}|}
\newcommand{\const}{{\rm const}} 
\newcommand{\atanh}{\,{\rm atanh}}

\newcommand{\ie}{i.e. }
\newcommand{\iet}{i.e.}
\newcommand{\eg}{e.g. }
\newcommand{\cc}{{\rm c.c.}} 
\newcommand{\hc}{{\rm h.c.}} 
\def\etal{{\it et al. }}

\newcommand{\jhatbf}{\hat {\textbf \j}} 
\newcommand{\Jhatbf}{\hat {\textbf \J}} 
\newcommand{\jhat}{\hat {\jmath}} 
\newcommand{\Jhat}{\hat {J}} 
\newcommand{\jbf}{\textbf j}
\newcommand{\Jbf}{\textbf J}

\def\chibf{\boldsymbol{\chi}}
\def\down{\downarrow}
\def\eps{\epsilon}
\def\gam{\gamma} 
\def\phibf{\boldsymbol{\phi}}
\def\varphibf{\boldsymbol{\varphi}}
\def\varphibfs{\boldsymbol{\varphi}_<}
\def\varphibfl{\boldsymbol{\varphi}_>}
\def\varphis{\varphi_{<}}
\def\varphil{\varphi_{>}}
\def\psibf{\boldsymbol{\psi}}
\def\Ome{\Omega}
\def\omeD{{\omega_D}} 
\def\bfOme{\boldsymbol{\Omega}} 
\def\Omebf{\boldsymbol{\Omega}} 
\def\lamb{\lambda}
\def\Lamb{\Lambda}
\def\sig{\sigma}
\def\Sig{\Sigma}
\def\sigp{{\sigma'}} 
\def\bfsig{\boldsymbol{\sigma}} 
\def\sigbf{\boldsymbol{\sigma}} 
\def\The{\Theta} 
\def\up{\uparrow}
\def\L{\Lambda}

\def\epsk{\epsilon_{\bf k}} 
\def\xik{\xi_{\bf k}} 
\def\xip{\xi_{\bf p}}
\def\xikq{\xi_{{\bf k}+{\bf q}}} 
\def\Ek{E_{\bf k}}
\def\Ep{E_{\bf p}}
\def\Heff{\hat H_{\rm eff}}
\def\Hem{\hat H_{\rm em}}
\def\Hint{\hat H_{\rm int}}
\def\Hloc{\hat H_{\rm loc}}
\def\HMF{\hat H_{\rm MF}}
\def\Sem{S_{\rm em}}
\def\SMF{S_{\rm MF}} 
\def\SRPA{S_{\rm RPA}} 
\def\Sint{S_{\rm int}} 
\def\Sloc{S_{\rm loc}} 
\def\Zloc{Z_{\rm loc}} 
\def\ZMF{Z_{\rm MF}} 
\def\ZRPA{Z_{\rm RPA}} 
\def\RPA{{\rm RPA}}
\def\loc{{\rm loc}} 
\def\pp{{\rm pp}}
\def\ph{{\rm ph}} 
\def\ch{{\rm ch}}
\def\sp{{\rm sp}} 
\def\qtf{q_{\rm TF}}
\def\epstf{\eps^{}_{\rm TF}} 
\def\epsrpa{\eps^{}_{\rm RPA}} 
\def\chinnzpp{\chi_{nn}^{0}{}\!\!\!''}

\def\half{\frac{1}{2}}
\def\dhalf{\dfrac{1}{2}}
\def\third{\frac{1}{3}} 
\def\quarter{\frac{1}{4}}

\def\qr{{\bf q}\cdot{\bf r}}
\def\wt{\omega t} 

\def\a{{\bf a}}
\def\b{{\bf b}}
\def\e{{\bf e}}
\def\f{{\bf f}}
\def\g{{\bf g}}
\def\h{{\bf h}}
\def\k{{\bf k}}
\def\l{{\bf l}}
\def\m{{\bf m}}
\def\n{{\bf n}} 
\def\p{{\bf p}} 
\def\q{{\bf q}}
\def\r{{\bf r}}
\def\t{{\bf t}}
\def\u{{\bf u}}
\def\v{{\bf v}}
\def\x{{\bf x}}
\def\y{{\bf y}} 
\def\z{{\bf z}} 
\def\A{{\bf A}}
\def\B{{\bf B}}
\def\D{{\bf D}} 
\def\E{{\bf E}} 
\def\F{{\bf F}} 
\def\H{{\bf H}}  
\def\J{{\bf J}}
\def\K{{\bf K}} 

\def\G{{\bf G}}
\def\M{{\bf M}}  
\def\O{{\bf O}} 
\def\P{{\bf P}} 
\def\Q{{\bf Q}} 
\def\R{{\bf R}}
\def\S{{\bf S}}
\def\epsbf{\boldsymbol{\epsilon}}
\def\mubf{\boldsymbol{\mu}}
\def\nablabf{\boldsymbol{\nabla}}
\def\rhobf{\boldsymbol{\rho}}
\def\sigmabf{\boldsymbol{\sigma}} 
\def\Pibf{\boldsymbol{\Pi}}
\def\pibf{\boldsymbol{\pi}}

\def\para{\parallel}
\def\kpara{{k_\parallel}}
\def\kperp{{k_\perp}} 
\def\kperpp{{k_\perp'}} 
\def\qperp{{q_\perp}} 
\def\tperp{{t_\perp}} 

\def\w{\omega}
\def\wn{\omega_n}
\def\wnu{\omega_\nu}
\def\wp{\omega_p} 
\def\dmu{{\partial_\mu}}
\def\dl{{\partial_l}}  
\def\dt{\partial_t} 
\def\tdt{\tilde\partial_t}
\def\dk{\partial_k}
\def\tdk{\tilde\partial_k}
\def\dx{\partial_x}
\def\dy{\partial_y} 
\def\dtau{{\partial_\tau}}  
\def\det{{\rm det}} 
\def\Pf{{\rm Pf}}

\def\dsum{\displaystyle \sum}
\def\dint{\displaystyle \int} 
\def\intt{\int_{-\infty}^\infty dt} 
\def\inttp{\int_{-\infty}^\infty dt'} 
\def\intk{\int_{\bf k}} 
\def\intkd{\int \frac{d^dk}{(2\pi)^d}}
\def\intq{\int_{\bf q}} 
\def\intr{\int d^dr}  
\def\dintr{\displaystyle \int d^dr} 
\def\intrp{\int d^dr'}
\def\dinttau{\displaystyle \int_0^\beta d\tau}
\def\dinttaup{\displaystyle \int_0^\beta d\tau'}
\def\inttau{\int_0^\beta d\tau}
\def\inttaup{\int_0^\beta d\tau'}
\def\intx{\int d^{d+1}x} 
\def\inttaur{\int_0^\beta d\tau \int d^dr}
\def\intinf{\int_{-\infty}^\infty}
\def\dinttaur{\displaystyle \int_0^\beta d\tau \int d^dr}
\def\dintinf{\displaystyle \int_{-\infty}^\infty}
\def\intw{\int_{-\infty}^\infty \frac{d\w}{2\pi}}
\def\sumr{\sum_{\bf r}} 

\def\calA{{\cal A}} 
\def\calC{{\cal C}} 
\def\dt{\partial_t}
\def\calD{{\cal D}}
\def\calF{{\cal F}} 
\def\calG{{\cal G}}
\def\calH{{\cal H}}
\def\calI{{\cal I}}
\def\calJ{{\cal J}}
\def\calK{{\cal K}}
\def\calL{{\cal L}} 
\def\calN{{\cal N}}
\def\calO{{\cal O}}
\def\calP{{\cal P}}  
\def\calR{{\cal R}} 
\def\calS{{\cal S}}
\def\calT{{\cal T}}
\def\calU{{\cal U}}
\def\calX{{\cal X}}
\def\calY{{\cal Y}} 
\def\calZ{{\cal Z}} 

\def\calFbf{{\bf F}}

\def\tT{{\tilde T}}
\def\talpha{{\tilde\alpha}}
\def\tdelta{{\tilde\delta}}
\def\teta{{\tilde\eta}} 
\def\tlamb{{\tilde\lambda}}
\def\tmu{{\tilde\mu}}
\def\tphibf{{\tilde\phibf}}
\def\trho{{\tilde\rho}}
\def\tvarphibf{{\tilde\varphibf}} 
\def\tw{{\tilde\omega}}
\def\twn{{\tilde\omega_n}}

\def\asinh{{\rm asinh}} 
\def\dw{\partial_\w}
\def\Gamloc{\Gamma_{\rm loc}}
\def\Vloc{V_{\rm loc}}
\def\Zloc{Z_{\rm loc}}
\def\Gn{G_{\rm n}}
\def\Gan{G_{\rm an}}
\def\nbarloc{\bar n_{\rm loc}} 
\def\Ibarll{\bar I_{\rm ll}}
\def\Ibartt{\bar I_{\rm tt}}
\def\Jbarllll{\bar J_{\rm ll,ll}}
\def\Jbartttt{\bar J_{\rm tt,tt}}
\def\Jbarltlt{\bar J_{\rm lt,lt}}
\def\Jbarlltt{\bar J_{\rm ll,tt}}
\def\Jbarttll{\bar J_{\rm tt,ll}}
\def\Jbarlllt{\bar J_{\rm ll,lt}}
\def\Jbarltll{\bar J_{\rm lt,ll}}
\def\Jbarttlt{\bar J_{\rm tt,lt}}
\def\Jbarlttt{\bar J_{\rm lt,tt}}
\def\gllb{\bar{G}_{{\rm ll}}}
\def\gttb{\bar{G}_{{\rm tt}}}
\def\gltb{\bar{G}_{{\rm lt}}}

%%%%%%%%%%%%%%%%%%%%%%%%%%%%%%%%%%%%%%%%%%%%%%
%AUTHOR'S MACRO
%%%%%%%%%%%%%%%%%%%%%%%%%%%%%%%%%%%%%%%%%%%%%%

\title{A Non-Perturbative Renormalization Group approach to quantum XY spin models}  

\author{A. Ran\c{c}on}
\affiliation{James Franck Institute and Department of Physics,
University of Chicago, Chicago, Illinois 60637, USA}

\begin{abstract}
We present a Lattice Non-Perturbative Renormalization Group (NPRG) approach to quantum XY spin models by using a mapping onto hardcore bosons. The NPRG takes as initial condition of the renormalization group 
flow the (local) limit of decoupled sites, allowing us to take into account the hardcore constraint exactly. The initial condition of the flow is equivalent to the large $S$ classical results of the corresponding spin system. Furthermore, the hardcore constraint is conserved along the RG flow, and 
 we can describe both local and long-distance fluctuations in a non-trivial way. We discuss a simple approximation scheme, and solve the corresponding flow equations. We compute both the zero-temperature thermodynamics and the finite temperature phase diagram on the square and cubic lattices. The NPRG allows us to recover the correct critical physics at finite temperature  in two and three dimensions. The results compare well with numerical simulations.
\end{abstract}

\maketitle

\section{Introduction}

The study of frustrated quantum spin systems is one of the great challenges in condensed matter. The intertwinement between quantum fluctuations and frustrated interactions is beyond mean-field  approaches and allows for the realization of exotic phases.\cite{MisguichBook} Numerical approaches usually suffer from the sign problem, are constraint to relatively small sizes or work well for quasi-one dimensional  systems. Numerous field theoretic approaches exist, which are based on mapping the quantum spin operators to either bosonic or fermionic operators, for instance using hardcore bosons,\cite{Matsubara1956} Schwinger bosons,\cite{Arovas1988}  or pseudo-fermions.\cite{Popov1988} While these methods have given useful insights into the physics of spin systems, they also suffer from important limitations. For example, these approaches artificially increase the size of the Hilbert space, and on-site constraints have to be implemented to project out unphysical states. These are however generally difficult to implement, and the projection of the unphysical states is usually done only on average, \emph{i.e.} at a mean-field level. (The same kind of mapping, and the mean-field  description of the associated constraints, is also used to study the Kondo physics.\cite{Coleman1984}) In the case of semi-classical (large $S$) approaches, the mean-field solutions, plus their spin-wave corrections, tend to underestimate quantum fluctuations, and break down when the system becomes disordered, as in a spin-liquid phase.

In this paper, we propose a Renormalization Group (RG) scheme to address quantum XY spin models using the mapping onto hardcore bosons.\cite{Matsubara1956} This approach treats the hardcore constraint exactly and is capable of describing both ordered and disordered states, and therefore might be interesting for the study of exotic phases of spin systems.  We note that a  fermionic renormalization group  approach has already been implemented for quantum spins using pseudo-fermions.\cite{Reuther2010,Reuther2011,Reuther2014} It can describe  different types of fluctuations on equal footing, and is thus useful to compute phase diagrams beyond mean-field theories. However, this approach is perturbative in the coupling constants, though functional in momentum and frequency, and is thus confined either to the description of disordered states or to large frequencies or momenta. Indeed, the presence of an ordered state is characterized by a divergence of the momentum-frequency dependent coupling constants, where the fermionic RG breaks down. In this respect, the approach described in this paper is well suited for describing ordered states, which will be represented by (potentially exotic) superfluid phases of the hardcore bosons.

Here we show how to implement a Lattice Non-Perturbative Renormalization Group (NPRG) technique for the simplest Hamiltonian for hardcore bosons (see Eq. \eqref{eq_H}), corresponding to the quantum XY model in a magnetic field. The strategy of the NPRG is to construct a family of models indexed by a momentum scale $k$, such that fluctuations are gradually included as one lowers $k$ from a microscopic scale $k=\Lambda$ (corresponding to an exactly solvable model) down to $k=0$ (where one recovers the model of interest).\cite{Berges2002,DelamotteIntro} This is done by adding a regulator term $\Delta \hat H_k$ to the Hamiltonian. The Lattice NPRG is characterized by its initial condition.\cite{Machado2010,Rancon2011,Rancon2011a} The regulator term is chosen such that at the microscopic scale $k=\Lambda$, the Hamiltonian $\hat H+\Delta\hat H_\Lambda$ corresponds to the limit of decoupled sites. The hardcore constraint is thus included right at the beginning of the RG flow. The inter-site coupling is then gradually restored as $k$ is lowered, and the hardcore constraint is conserved along the flow. At small $k\ll \Lambda$, $\Delta \hat H_k$ plays the role of an infrared regulator that suppresses the long-distance fluctuations, and the RG flow is equivalent to that of the standard NPRG, thus giving the same results for critical properties close to a phase transition. The Lattice NPRG has proven to be a successful method for computing both universal and non-universal quantities, such as the thermodynamics and phase diagrams, for classical and quantum systems.\cite{Machado2010,Rancon2011,Rancon2011a,Rancon2012,Rancon2012a,Rancon2012b,Rancon2013a,Taranto2014}

This paper is organized as follows. In Sec. \ref{sec_LNPRG} we describe the Lattice NPRG for hardcore bosons. We introduce a scale-dependent effective action and discuss the initial condition of the RG flow. We show that it is given by the classical solution of the corresponding spin system. We discuss the flow equations and present a simple approximation scheme. This section closely follows that of Ref. \onlinecite{Rancon2011a}, but will allow for a clear and self-contained introduction to the approach.
In Sec. \ref{sec_thermo}, we compute the thermodynamics  at zero and finite temperature in two and three dimensions, and find a good agreement with recent Monte Carlo simulations.  We show that the thermodynamics for densities close to zero or one is well described in terms of universal functions given by Bogoliubov theory, corresponding to the limit of dilute particles or holes, respectively. We recover the correct finite temperature physics, in particular the critical regime close to the critical temperature in dimension three, as well as the Berezinskii-Kosterlitz-Thouless (BKT) physics in dimension two. The main results are summarized in the conclusion (Sec. \ref{sec_concl}). Some additional technical details are given in Appendix \ref{app_Gam}, \ref{app_temp}, and \ref{app_highw}, and we show that the NPRG can recover the harmonic spin-wave corrections exactly in Appendix \ref{app_1loop}.

\section{Lattice NPRG \label{sec_LNPRG}}

The simplest model of XY quantum spins-$\half$ is given by the Hamiltonian
\begin{equation}
\hat H_{\rm XY} = -2t\sum_{\langle\r,\r'\rangle}\big[\hat \sigma^x_\r\, \hat \sigma^x_{\r'}+\hat \sigma^y_\r\, \hat \sigma^y_{\r'}\big]-\mu\sum_\r \big[ \hat \sigma^z_\r +1/2\big],
\label{eq_HXY}
\end{equation}
where $t$ is the spin interaction energy in the $x$-$y$ plane between nearest-neighbor sites $\langle\r,\r'\rangle$,  and $\mu$ is a constant magnetic field along the $z$-axis (up to a constant introduced for later convenience). Here $\sigma^a_\r$ are spin-$1/2$ operators, which satisfy the $SU(2)$ commutation relations $ \big[\hat \sigma^a_\r,\hat \sigma^b_{\r'}\big]=\delta_{\r,\r'}i\epsilon^{abc}\hat \sigma^c_\r$. This model has been studied in detail using a number of methods.\cite{Matsubara1956,Hamer1994,Micnas1995,Pedersen1996,Bernardet2002,Melko2005,Coletta2012,Carrasquilla2012}

Following Matsubara and Matsuda,\cite{Matsubara1956} we can map the spin operators onto hardcore bosons using $\hat b_\r= \hat \sigma^x_\r-i\hat \sigma^y_\r$, $\hat b^\dag_\r= \hat \sigma^x_\r+i\hat \sigma^y_\r$ and $\hat b^\dag_\r\hat b_\r=\sigma^z_\r +1/2$, where the eigenstates $\ket\down$ and $\ket\up$ of $\hat \sigma^z$ are mapped onto the states with zero and one boson, respectively. One can check that the $SU(2)$ algebra is recovered if the creation ($\hat b^\dag_\r$) and annihilation ($\hat b_\r$)  operators respect the hardcore bosonic commutation relation $\big[\hat b_\r\,, \hat b^\dag_{\r'}\big]=\delta_{\r,\r'}(1-2 \hat b^\dag_\r \,\hat b_\r)$ in addition to the hardcore constraints $\big(\hat b_\r^{(\dag)}\big)^2=0$.

The spin Hamiltonian is thus rewritten as
\begin{equation}
\hat H = -t\sum_{\langle\r,\r'\rangle}\big[\hat b^\dag_\r\, \hat b_{\r'}+\hat b^\dag_{\r'}\, \hat b_{\r}\big]-\mu\sum_\r  \hat b^\dag_\r \,\hat b_\r\,,
\label{eq_H}
\end{equation}
which is the Hamiltonian we will study in the rest of the paper.  Here, $t$ is interpreted as a hopping amplitude between neighboring sites  and $\mu$ as a chemical potential for the bosons. Note that spin-spin interactions along the $z$-axis $\hat \sigma^z_\r\, \hat \sigma^z_{\r'}$, that appear for example in the Heisenberg model, would correspond to interactions between neighboring sites for the hardcore bosons. These interactions are not straightforward to implement in the current version of the Lattice NPRG, but we briefly comment on how to circumvent this problem in the conclusion.

We set $\hbar=k_B=1$, as well as the lattice spacing as unit length throughout the paper.

\subsection{Scale-dependent effective action}
Following the general strategy of the Lattice NPRG, we consider a family of models with Hamiltonian $\hat H_k=\hat H+\Delta \hat H_k$ indexed by a momentum scale $k$ varying from a microscopic scale $\Lambda$ down to 0. The regulator term is defined by  
\begin{equation}
\Delta \hat H_k=\sum_\q  R_k(\q)  \,\hat b_\q^{\dag}\,\hat b_\q\,,
\label{eq_Hreg}
\end{equation}
where $\hat b_\q$ is the Fourier transform of $\hat b_\r$ and the sum over $\q$ runs over the first Brillouin zone $]-\pi,\pi]^d$ of the reciprocal lattice. The cutoff function $R_k(\q)$ modifies the bare dispersion $t_\q=-2t\sum_{i=1}^d \cos q_i$ of the bosons (from now on, we assume that the bosons are on a hypercubic lattice, though most of the discussion of this section is more general). $R_\Lambda(\q)$ is chosen such that the effective (bare) dispersion $t_\q+R_\Lambda(\q)$ vanishes.\cite{Machado2010,Rancon2011,Rancon2011a} The Hamiltonian $\hat H_\Lambda=\hat H+\Delta \hat H_\Lambda$ then corresponds to the local limit of decoupled sites (vanishing hopping amplitude). With the choice $R_\Lambda(\q)+t_\q=0$, only the hopping is modified and the local Hamiltonian is unchanged. 

Here, we use the cutoff function ~\cite{Rancon2011,Rancon2011a}
\begin{equation} 
R_k(\q) = - Z_{A,k} \eps_k \mbox{sgn}(t_\q) (1-y_\q)\Theta(1-y_\q) \,,
\label{cutoff} 
\end{equation}
with $\Lambda=\sqrt{2d}$, $\eps_k=tk^2$, $y_\q=(2dt-|t_\q|)/\eps_k$ and $\Theta(x)$ the step function. The $k$-dependent constant $Z_{A,k}$ is defined below ($Z_{A,\Lambda}=1$). 
Since $R_{k=0}(\q)=0$, the Hamiltonian $\hat H_{k=0}$ coincides with the Hamiltonian~(\ref{eq_H}) of the original model. For small $k$, the function $R_k(\q)$ gives a mass $\sim k^2$ to the low-energy modes $|\q|\lesssim k$ and acts as an infrared regulator as in the standard NPRG scheme.\cite{Berges2002,DelamotteIntro} 

The physics of the system is completely described by the source-dependent partition function
\begin{equation}
Z_k[J,J^*]=\Tr\left\{ T_\tau e^{-\int_0^\beta d\tau\,\left(\hat H_k-\sum_\r\left[ J_\r(\tau)\hat b_\r^\dag(\tau)+ J_\r^*(\tau)\hat b_\r(\tau)\right]\right) }       \right\}\,,
\end{equation}
where $\beta=1/T$ is the inverse temperature,  $T_\tau$ is the imaginary time ordering operator and $\hat b_\r^{(\dag)}(\tau)=e^{\tau\hat H}b_\r^{(\dag)}e^{-\tau\hat H}$. If evaluated at vanishing sources $J=J^*=0$, one recovers the (scale-dependent) partition function, from which one obtains the thermodynamics.\footnote{Finite (and time-independent) sources correspond in the quantum spin language to a magnetic field in the $x$-$y$ plane.} Furthermore, functional derivatives with respect to the sources give access to all correlation functions. In particular, the (source-dependent) superfluid order parameter $\phi^{(*)}_\r(\tau)=\mean{\hat b_\r^{(\dag)}(\tau)}$ is given by 
\begin{equation}
\phi_\r(\tau)=\frac{\delta \ln Z_k[J,J^*]}{\delta J^*_\r(\tau)},  \,\,\, \phi^*_\r(\tau)=\frac{\delta \ln Z_k[J,J^*]}{\delta J_\r(\tau)}\,.
\end{equation}
We can then introduce the scale-dependent effective action 
\begin{align}
\Gamma_k[\phi^*,\phi] ={}& - \ln Z_k[J^*,J] + \inttau \sum_\r (J^*_\r\phi_\r+\mbox{c.c.}) \nonumber \\ & - \Delta H_k[\phi^*,\phi], 
\end{align}
defined as a (modified) Legendre transform of $\ln Z_k[J,J^*]$ which includes the explicit subtraction of $\Delta H_k[\phi^*,\phi]=\int_0^\beta d\tau\sum_\q  R_k(\q)  \, \phi^*_\q(\tau)\,\phi_\q(\tau)$. By varying the scale $k$, a renormalization group equation can be obtained for the scale-dependent effective action,\cite{Wetterich1993} 
\begin{equation}
\partial_k \Gamma_k[\phi^*,\phi] = \half \Tr\biggl\lbrace \partial_k R_k\left(\Gamma^{(2)}_k[\phi^*,\phi] + R_k\right)^{-1} \biggr\rbrace \,,
\label{wetteq}
\end{equation}
where $\Gamma^{(2)}_k$ is the second-order functional derivative of $\Gamma_k$. In Fourier space, the trace in (\ref{wetteq}) involves a sum over momenta and frequencies as well as the two components of the complex field $\phi$. 
\footnote{Note that we do not need to use a functional integral formalism to obtain a RG equation for the effective action. In particular, such a functional integral would be difficult to write down, since hardcore bosons do not obey canonical (bosonic or fermionic) commutation relations.}

 For the purpose of this work, we can concentrate on two quantities, the first one being the effective potential
\begin{equation}
V_k(n)=\frac{1}{\beta N} \Gamma_k[\phi,\phi^*]\big|_{\phi\,\, {\rm const}}\,,
\label{eq_effpot}
\end{equation}
where $\phi$ is a uniform and time-independent field, and $N$ is the number of lattice sites. The global $U(1)$ symmetry of the Hamiltonian $\hat b^{(\dag)}_\r\to e^{\pm i\alpha} \hat b^{(\dag)}_\r$ implies that $V_k(n)$ is a function of $n = |\phi|^2$ (not to be confused with the density of particle $\bar n$). Its minimum determines the
condensate density $n_{0,k}$ and the thermodynamic potential (per site) $V_{0,k}= V_k(n_{0,k})=-P_k$ in the equilibrium state, where $P_k$ is the (scale-dependent) pressure.

The second quantity of interest is the two-point vertex function
\begin{equation}
\Gamma^{(2)}_{k,ij}(\r-\r',\tau-\tau';\phi) = \frac{\delta^{(2)} \Gamma_k[\phi]}{\delta\phi_{i\r}(\tau) \delta\phi_{j\r'}(\tau')} \biggl|_{\phi\,\,\const} , 
\label{eq_defG}
\end{equation}
which determines the one-particle propagator $G_k=-\Gamma_k^{(2)-1}$. Here the indices $i,j$ refer to the real and imaginary parts of $\phi=\frac{1}{\sqrt{2}}(\phi_1+i\phi_2)$. Because of the U(1) symmetry of the Hamiltonian, the two-point vertex function in a constant field takes the form ~\cite{Dupuis2009a} 
\begin{equation}
\Gamma_{k,ij}^{(2)}(q;\phi) = \delta_{i,j}\,\Gamma_{A,k}(q;n) + \phi_i\phi_j\, \Gamma_{B,k}(q;n) + \eps_{ij}\, \Gamma_{C,k}(q;n) 
\label{gam2}
\end{equation}
in Fourier space, where $q=(\q,i\w)$, $\w$ is a Matsubara frequency and $\eps_{ij}$ the antisymmetric tensor. The symmetries of the two-point function, and an explicit expression for the propagator are given in Appendix \ref{app_Gam}.

\subsection{Initial conditions \label{subsec_initial}}

\subsubsection{Local effective action}
Since the Hamiltonian $\hat H_\Lambda=\hat H+\Delta\hat H_\Lambda$ corresponds to the local limit, the initial value of the scale-dependent effective action is given by
\begin{equation}
\Gamma_\Lambda[\phi^*,\phi] = \Gamma_\loc[\phi^*,\phi] + \inttau \sum_\q \phi^*_\q(\tau) t_\q \phi_\q(\tau)\, ,
\label{GamLambda}
\end{equation}
where 
\begin{equation}
\Gamma_\loc[\phi^*,\phi] = -\ln Z_\loc[J^*,J] + \inttau \sum_\r (J^*_\r\phi_\r+\mbox{c.c.}) 
\label{Gamloc}
\end{equation}
is the Legendre transform of the thermodynamic potential $-\ln \Zloc[J^*,J]$ in the local limit. In Eq.~(\ref{Gamloc}), $J$ is related to $\phi$ by the relation $\phi_\r(\tau)=\delta \ln \Zloc[J^*,J]/\delta J^*_\r(\tau)$ and $\Zloc$ is the partition function obtained from  $\hat H_\Lambda$. 

Even for this simple Hamiltonian, it is not possible to compute the functional $\Gamma_\loc[\phi^*,\phi]$ for arbitrary time-dependent fields. However the effective potential $V_\loc(n)$ and the two-point vertex function $\Gamma_\loc^{(2)}$ are easily calculated in a time-independent field $\phi$, which are the two quantities we will need, given the approximations made on the flow equations (Sec.~\ref{subsec_rgeq}).

\begin{figure}[t]
\centerline{\includegraphics[width=6cm]{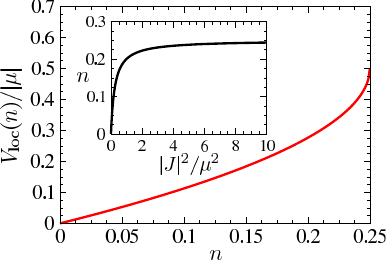}}
\caption{(Color online) Local effective potential $V_\loc(n)$ as a function of $n$ for a negative chemical potential (for $\mu>0$, the potential is shifted by $-\mu$). Inset: $n=|\phi|^2$ as a function of the source $|J|^2$.  } 
\label{fig_Vloc} 
\end{figure}

To calculate $V_\loc$ and $\Gamma_\loc^{(2)}$ in a constant field, it is sufficient to solve the problem of a single site with constant complex external source $J$. The corresponding Hamiltonian reads
\begin{equation}
\hat H_\loc(J,J^*) = -\mu \,\hat b^\dag \hat b - J^* \hat b - J \hat b^\dagger \,,
\label{eq_Hloc}
\end{equation}
and is trivially diagonalized. Labeling the eigenvectors  $\ket \pm$ and the corresponding eigenvalues $E_\pm(J,J^*)$ (with $\ket -$ the groundstate), we obtain
\begin{equation}
E_\pm(J,J^*)=-\frac{\mu}{2}\pm \frac{\sqrt{4|J|^2+\mu^2}}{2}\,.
\end{equation}
At zero temperature $\beta\to\infty$ (the finite temperature case is discussed in Appendix \ref{app_temp}), we obtain the  superfluid order parameter
\begin{equation}
\begin{split}
\phi^{(*)} &= \frac{J^{(*)}}{\sqrt{4|J|^2+\mu^2}}\,,\\
|\phi|^2 &=\frac{|J|^2}{4|J|^2+\mu^2}\,,
\end{split}
\label{phiJ}
\end{equation}
and the effective potential
\begin{equation}
\begin{split}
V_\loc(n) &= E_- + J^*\phi + J \phi^* \,,\\
&=-\frac{\mu}{2}-\frac{|\mu|}{2}\sqrt{1-4n}\,,
\end{split}
\label{eq_Vloc}
\end{equation} 
with $n=|\phi|^2$. Figure~\ref{fig_Vloc} shows the superfluid order parameter $\phi$ as a function of the external source $J$, and the local effective potential $V_{\rm loc}(n)$. Note that due to the hardcore constraint, the field is bounded $0\leq n \leq \frac{1}{4}$. This can be understood easily with the following reasoning: any (one-site) state can be written as $\ket{\psi}=\cos\theta \ket{0}+e^{i\Phi}\sin{\theta}\ket{1}$, with $\theta$ and $\Phi$ two angles ($\ket{0}$ and $\ket{1}$ are the state with zero and one boson, respectively). From this we get $\phi=\mean{\hat b}=e^{i\Phi}\cos\theta\sin\theta =e^{i\Phi}\sin(2\theta) /2$, hence $n=|\phi|^2$ cannot be larger than $1/4$.\footnote{We remind the reader that $n$ is not the density of particles, which is denoted by $\bar n$.} Although $V_\loc(n)$ is finite for all $n\in[0,\frac{1}{4}]$, its derivatives are diverging at $n= \frac{1}{4}$, as can be seen from
\begin{equation}
\frac{\partial^i V_\loc(n)}{\partial n^i} =A_i |\mu|(1-4n)^{\half-i}\,,
\end{equation}
where $A_{i+1}=2(2i-1)A_i$ and $A_0=-\half$. This recurrence equation can be solved and yields $A_i=4^{i-1}\Gamma(i-\half)/\Gamma(\half)$, where $\Gamma(z)$ is the Gamma function ($A_i=\frac{(2i-2)!}{(i-1)!}$ if $i\geq1$).

The fact that the derivatives of $V_\loc$ diverge can be understood as follow: because $n$ cannot be larger than $\frac{1}{4}$, quantum fluctuations make it difficult to increase $n$ to its maximum. Stated otherwise, it costs a lot of energy to increase $n$, and an infinite amount to have $n=\frac{1}{4}$, rendering it impossible to get $n>\frac{1}{4}$. Furthermore, this property is preserved along the RG flow, since Eq. \eqref{wetteq} contains the derivatives of the effective action in the denominator, implying that fluctuations that try to increase the value of the order parameter close to $\frac{1}{4}$ will be suppressed. The boundedness of the order parameter in the context of the Lattice NPRG has already been observed in the case of classical spin systems, where the magnetization cannot be greater than one. \cite{Machado2010} 

Remark that $V_\loc(n)$ is not well defined for $\mu=0$ as the field becomes independent of the source ($n=\frac{1}{4}$ for all $|J|$). This can be traced back to the fact that $\hat H_\loc(0,0)=0$ in this limit and that the bosons do not have any dynamics. However, we will see that all physical quantities computed with the Lattice NPRG are well defined in the limit $\mu\to 0$.

To determine the two-point vertex $\Gamma^{(2)}_{\rm loc}$, we start from the (source-dependent) normal and anomalous local Green functions
\begin{equation}
\begin{split}
G_{\rm n}(\tau) &= - \mean{T_\tau \hat b(\tau) \hat b^\dagger(0)} + |\mean{\hat b}|^2 \,, \\ 
G_{\rm an}(\tau) &= - \mean{T_\tau \hat b(\tau) \hat b(0)} + \mean{\hat b}^2 \,.
\label{eq_Green}
\end{split}
\end{equation}
 The Fourier transforms $G_{\rm n}(i\w)$ and $G_{\rm an}(i\w)$ are easily expressed in terms of the eigenstates $\ket{\pm}$ of the Hamiltonian,
\begin{equation}
\begin{split}
G_{\rm n}(i\w) &= - \left[ \frac{|\bra{+} \hat b \ket{-}|^2}{i\w+E_+-E_-} - \frac{|\bra{-} \hat b \ket{+}|^2}{i\w+E_--E_+} \right]\, , \\ 
G_{\rm an}(i\w) &= -  \bra{+} \hat b \ket{-}\bra{-} \hat b \ket{+}  \frac{2(E_+-E_-)}{\w^2+(E_+-E_-)^2}\,  .
\end{split}
\end{equation}
From the relation $\Gamma^{(2)}=-G^{-1}$, we obtain
\begin{equation}
\begin{split}
\Gamma_{A,\loc}(i\w;n) &= -\frac{1}{2D} [G_{\rm n}(i\w)+G_{\rm n}(-i\w)+2G_{\rm an}(i\w)] \,, \\ 
\Gamma_{B,\loc}(i\w;n) &= \frac{G_{\rm an}(i\w)}{nD} , \\ 
\Gamma_{C, \loc}(i\w;n) &= \frac{i}{2D} [G_{\rm n}(i\w)-G_{\rm n}(-i\w)] \,, 
\end{split}
\label{eq_GamG}
\end{equation}
where $D=G_{\rm n}(i\w)G_{\rm n}(-i\w)-G_{\rm an}(i\w)^2$. $\Gamma_\loc^{(2)}$ is expressed in terms of the condensate density $n$ (rather than the external source $J$) by inverting (\ref{phiJ}). 

An explicit calculation gives
\begin{equation}
\begin{split}
\Gamma_{A,{\rm loc}}(i\w;n) &= V'_\loc(n)\, , \\ 
\Gamma_{B,\loc}(i\w;n) &= V''_\loc(n) \,, \\ 
\Gamma_{C,\loc}(i\w;n) &= Z_{C,\loc}(n)\, \w= -\frac{{\rm sgn}(\mu)}{\sqrt{1-4n}}\w\, , 
\end{split}
\label{eq_G2loc}
\end{equation}
showing that $\Gamma_{A,\loc}$ and $\Gamma_{B,\loc}$ are frequency independent, whereas $\Gamma_{C,\loc}$ is linear in frequency. Furthermore the sign of $Z_C$ reflects the particle- (hole-) type of the local excitations for negative (positive) chemical potential. The results of Eq. \eqref{eq_G2loc} are in agreement with the (zero-temperature) Ward identities ~\cite{Dupuis2009a,Rancon2011a}
\begin{equation}
\begin{split} 
\frac{\partial}{\partial\w} \Gamma_{C,k}(q;n) \biggl|_{q=0} &=  - \frac{\partial^2 V_k(n)}{\partial n\, \partial\mu}\, , \\ 
\frac{\partial^2}{\partial\w^2} \Gamma_{A,k}(q;n) \biggl|_{q=0} &=  - \frac{1}{2n} \frac{\partial^2 V_k(n)}{\partial\mu^2} \,.
\end{split}
\label{wardid}
\end{equation}
The results of Eq. \eqref{eq_G2loc} also agree with the high frequency behavior of the local Green functions, see Appendix \ref{app_highw}.

\subsubsection{Mean-field solution }
The initial effective action $\Gamma_\Lambda$ [Eq.~(\ref{GamLambda})] treats the local fluctuations exactly but includes the inter-site hopping term at the mean-field level, reproducing the classical (large $S$) solution  of the XY quantum spin model.\cite{Matsubara1956,Bernardet2002,Coletta2012} Alternatively, we can think of  $\Gamma_\Lambda$ as the effective action of the equivalent of the strong-coupling Random Phase Approximation (RPA) theory used in the context of the Bose-Hubbard model. \cite{Sengupta2005} 

The effective potential reads
\begin{equation}
V_\Lambda(n) = V_\loc(n) - 2dtn\, , 
\end{equation}
while the two-point vertex function takes the RPA-like form 
\begin{equation}
\Gamma^{(2)}_{\Lambda,ij}(q;n) = \Gamma_{{\rm loc},ij}^{(2)}(i\w;n) + \delta_{i,j}t_\q \,. 
\end{equation}
Expanding $V_\Lambda(n)$ about $n=0$, we find 
\begin{equation}
V_\Lambda(n) = -\frac{\mu+|\mu|}{2} +  \left(|\mu|-2dt \right) n + \calO(n^2) \,, 
\label{Vexpand} 
\end{equation}
The ground state is disordered as long as $V_\Lambda'(0)\geq 0$. Thus the transition to the superfluid state is determined by the criterion $V'_\Lambda(0)=0$, \emph{i.e.}
\begin{equation}
\mu=\mu_c^\pm=\pm 2dt\,. 
\label{mfcrit}
\end{equation}
Equation~(\ref{mfcrit}) can also be obtained from the condition $\det\,\Gamma^{(2)}_\Lambda(\q=i\w=0;n=0)$, which signals the appearance of a pole at zero momentum and frequency in the one-particle propagator $G_\Lambda=-\Gamma^{(2)-1}_\Lambda$. In fact, it is well known that Eq. \eqref{mfcrit} is exact: for $\mu\leq -2dt$ ($\mu\geq 2dt$) the system is empty (full);  because the dynamics is non-relativistic, the propagator cannot be renormalized, since there are no other particles (holes) with which to interact. Therefore the critical chemical potential is unchanged when fluctuations are included.

The condensate density $n_{0,\Lambda}$ in the superfluid phase is determined by
\begin{equation}
V_\Lambda'(n_{0,\Lambda}) = V_\loc'(n_{0,\Lambda}) - 2dt = 0\,. 
\end{equation}
The hopping amplitude $t$ acts as a source term for the local potential $V_\loc(n)$. This gives
\begin{equation}
n_{0,\Lambda}=\frac{1}{4}(1-\bar \mu^2)\,,
\end{equation}
where we have introduced $\bar \mu=\mu/(2dt)$. The pressure is given by 
\begin{equation}
P_\Lambda=-V_\Lambda(n_{0,\Lambda})=\frac{dt}{2}(1+\bar\mu)^2\,,
\end{equation}
and the density $\bar n=\partial P/\partial\mu$ reads
\begin{equation}
\bar n_\Lambda=\half (1+\bar \mu)\,.
\end{equation}
We can also obtain the dispersion relation in the superfluid phase using $\det\,\Gamma^{(2)}_\Lambda(q;n_\Lambda)=0$ (after analytic continuation $i\w\to \w+i0^+$)
\begin{equation}
E_{\q}=\sqrt{\epsilon_\q\big(\bar\mu^2\epsilon_\q+2dt(1-\bar\mu^2)\big)}\,,
\label{eq_dispMF}
\end{equation}
where we have introduced $\epsilon_\q=t_\q+2dt$.
All these results are in perfect agreement with the large $S$ (mean-field) calculations. In fact, the NPRG being one-loop exact, one can recover exactly the harmonic (spin-wave) corrections, as shown in Appendix \ref{app_1loop}. However, in the context of hardcore bosons, there is no obvious small parameter that would allow us to control a perturbative expansion.

\subsection{Approximation of the RG equation}
\label{subsec_rgeq} 

To obtain the flow equations, we follow here the same reasoning that was used in Refs. \onlinecite{Rancon2011,Rancon2011a} and that we briefly recall.
Because one cannot obtain the explicit form of the scale-dependent effective action $\Gamma_k[\phi^*,\phi]$ at $k=\Lambda$, we cannot write an ansatz for it that could be used in combination with the flow equation \eqref{wetteq}, as it is usually done in the standard implementation of the NPRG.\cite{Berges2002,DelamotteIntro} However, as we have seen above, we can compute both the effective potential $V_k(n)$ and the two-point vertex function in constant field $\Gamma^{(2)}_k(q;n)$ which are the quantities of interest in the  Blaizot--M\'endez-Galain--Wschebor (BMW) scheme.\cite{Blaizot2006,Benitez2009,Benitez2012}  Therefore, the BMW approximation that we implement here allows us to obtain closed RG equations for both quantities. An additional approximation is to use a derivative expansion of the two-point vertex function to obtain the propagators that are needed in the evaluation of the flow equations.  This is always possible, as the function $R_k(\q)$ acts as an infrared regulator and $\Gamma^{(2)}_k(q;n)$ is regular in $q$ for $q\to 0$, and we approximate it by
\begin{equation}
\begin{split}
\Gamma_{A,k}(q;n) &= Z_{A,k}(n)\eps_\q+ V_{A,k}(n) \w^2 + V_k'(n)\, , \\ 
\Gamma_{B,k}(q;n) &= V''_k(n)\,, \\ 
\Gamma_{C,k}(q;n) &= Z_{C,k}(n)\w\, , 
\end{split}
\label{gamde}
\end{equation} 
in agreement with the symmetries of the two-point vertex function (see Eq. \eqref{sym}). Note that even though $V_{A,k}(n)$ is initially zero, it is important to include it in order to properly describe the infrared behavior in the superfluid phase. \cite{Dupuis2009,Dupuis2009a,Rancon2011,Rancon2011a}
Furthermore, it is crucial to keep the full lattice structure in the early stages of the RG flow ($k\simeq\Lambda$), and we have kept the full dispersion in $\Gamma_{A,k}$. Following Ref.~\onlinecite{Machado2010}, $Z_{A,k}(n)$ is defined as 
\begin{equation}
Z_{A,k}(n) = \frac{1}{t} \lim_{q\to 0} \frac{\partial}{\partial \q^2} \Gamma_{A,k}(q;n)\, , 
\label{ZAdef}
\end{equation}
so that $Z_{A,k}(n_{0,k})$ is a field renormalization factor.~\cite{Machado2010,Dupuis2008} 

Solving the flow equations for the functions $V_k(n)$, $Z_{C,k}(n)$,  $V_{A,k}(n)$ and $Z_{A,k}(n)$ is not a simple numerical task. Indeed, at zero temperature, the system is always in the ordered phase (for the non-trivial case $|\mu|<2dt$) implying that the effective potential becomes a convex function during the flow: the potential must be flat between $n=0$ and $n=n_{0,k=0}$ ($n_{0,k}$ determines the position of the minimum of $V_k(n)$, see below). Furthermore, the hardcore constraint implies that all derivatives of these functions will be singular as $n$ goes to $\frac{1}{4}$. Moreover, we have shown in Appendix \ref{app_1loop} that the NPRG reproduces exactly, in a loop expansion, the spin-wave corrections to the mean-field result, which already compare very well with the Monte Carlo simulations on the square lattice.\cite{Coletta2012} This implies that the NPRG has the capability to do as well as the large $S$ expansion at the level of the thermodynamics. 

We therefore use in the rest of the paper the simplest approximations described now, which will also compare well with the  Monte Carlo calculations. This will still show the power of the method while allowing us to concentrate on the physics. The numerical solution of the flow equations can be further simplified by expanding $V_k(n)$, $Z_{C,k}(n)$,  $V_{A,k}(n)$ and $Z_{A,k}(n)$ around $n_{0,k}$. Because $V_{A,k}(n)$ and $Z_{A,k}(n)$ are field independent at $k=\Lambda$, approximating them by $Z_{A,k}\equiv Z_{A,k}(n_{0,k})$ and $V_{A,k}\equiv V_{A,k}(n_{0,k})$ should be a good approximation (it is one loop exact). The field dependence of $V_k(n)$ and $Z_{C,k}(n)$ is more important, but we will nevertheless  approximate $Z_{C,k}(n)$ by $Z_{C,k}\equiv Z_{C,k}(n_{0,k})$, and expand the effective potential to quadratic order about its minimum, 
\begin{equation}
V_k(n) = \left\lbrace 
\begin{array}{lcc}
V_{0,k} + \frac{\lambda_k}{2}(n-n_{0,k})^2 & \mbox{if} & n_{0,k}>0 \,, \\[0.2cm] 
V_{0,k} + \delta_k n + \frac{\lambda_k}{2}n^2 & \mbox{if} & n_{0,k}=0 \,,
\end{array}
\right. 
\label{trunc} 
\end{equation}
where the condensate density $n_{0,k}$ is defined by
\begin{equation}
\frac{\partial V_k(n,\mu)}{\partial n}\biggl|_{n_{0,k}} = 0\,. 
\label{n0def}
\end{equation}
 One can systematically improve the NPRG results by increasing the order of the expansion. 
 %For example, to obtain a one-loop exact condensate density, one would have to expand the potential to order $(n-n_{0,k})^3$ and $Z_{C,k}(n)$ to order  $(n-n_{0,k})^1$.
Nevertheless, these approximations have been shown to be very successful in describing the thermodynamics of bosons in the continuum and on the lattice in two and three dimensions, either in the dilute regime or close to the Mott transition, at zero or finite temperature. \cite{Floerchinger2009,Rancon2011,Rancon2011a,Rancon2012,Rancon2012a,Rancon2012b,Rancon2013a}
The approximated flow equations  are detailed in Appendix~D of Ref. \onlinecite{Rancon2011a}, see in particular Eqs. (D4), (D5) and (D6).

\section{Thermodynamics \label{sec_thermo}}
We now discuss the thermodynamics and the finite temperature phase diagram in two and three dimensions. To do so, we solve the flow equations, with the approximations discussed above for a given value of $\mu$, $t$, and $\beta$. To simplify the notations, the subscript $k$ is dropped
whenever we refer to a $k = 0$ quantity (\emph{e.g.} $n_0 \equiv n_{0,k=0}$).

\subsection{Zero temperature}

\begin{figure}
\centerline{\includegraphics[width=6cm]{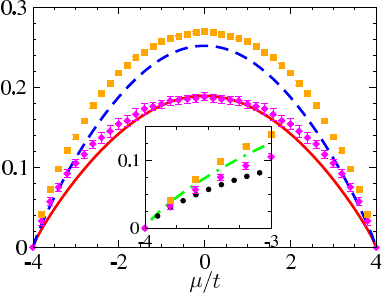}}
\caption{(Color online) Condensate density $n_0$ (solid line) and superfluid density $n_s$ (dashed line) as a function of $\mu$ at $T=0$ in two dimensions. Symbols show the Monte Carlo calculations of Ref. \onlinecite{Coletta2012} for $n_0$ (diamonds) and $n_s$ (squares).  Inset: Behavior close to the critical point $\mu_c=-4t$. The dotted line and the dot-dashed line correspond  to the dilute limit of $n_0$ and $n_s$ respectively. For clarity, the NPRG calculation is not shown, but it fails to correctly reproduce the dilute limit.   } 
\label{fig_T02D} 
\end{figure}

\begin{figure}
\centerline{\includegraphics[width=6cm]{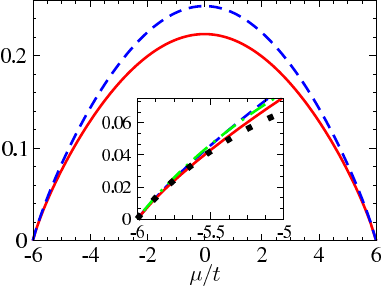}}
\caption{(Color online) Condensate density $n_0$ (solid line) and superfluid density $n_s$ (dashed line) as a function of $\mu$ at $T=0$ in three dimensions. Inset: Behavior close to the critical point $\mu_c=-6t$. The dotted line and the dot-dashed line correspond  to the dilute limit of $n_0$ and $n_s$ respectively.  } 
\label{fig_T03D} 
\end{figure}

Figures \ref{fig_T02D} and  \ref{fig_T03D} show the zero-temperature condensate density $n_0$ and superfluid density, defined as $n_{s} =Z_{A}(n_{0})n_{0}$,~\cite{Dupuis2009}  in two and three dimensions, respectively. Notice that even if the initial conditions are singular in the limit $\mu\to0$ (Sec. \ref{subsec_initial}), all physical quantities are well defined in this limit (see also Appendix \ref{app_1loop}). Even with the simple approximations made here, the results compare well with the Monte Carlo calculations in two dimensions.\cite{Coletta2012} There are much fewer numerical calculations for ground-state properties in three dimensions (but see for example Refs. \onlinecite{Hamer1994,Micnas1995} for spin-wave calculations), but we can compare our calculation at half-filling $\mu=0$ to the recent simulations of Ref. \onlinecite{Melko2005} for the ground-state energy and the superfluid density, see Table \ref{table_3D}. Our calculations differ by a few percent from the Monte Carlo. %\footnote{In both two and three dimensions, the spin-wave corrections account for almost all the corrections to the mean-field results. Had we not truncated both $Z_{C,k}(n)$ and $V_k(n)$, we would have obtained the same results, as shown in appendix \ref{app_1loop}. } 
\begin{table}[h!]
\renewcommand{\arraystretch}{1.5}
\begin{center}
\begin{tabular}{ccc}
\hline \hline
 & NPRG & MC \\
\hline
$P$ & 1.576 & 1.58364(4) \\
\hline
$n_s$ & 0.254 & 0.2623(2)  \\
\hline \hline
\end{tabular}
\end{center}
\caption{Zero temperature pressure $P$ and superfluid density $n_s$ on a cubic lattice for $\mu=0$. Monte Carlo (MC) data from Ref. \onlinecite{Melko2005}.}
\label{table_3D}
\end{table}

The thermodynamics close to the points $\mu=\mu^-_c=-2dt$ can be interpreted using the standard physics of the dilute Bose gas due to the presence of a quantum critical point. Indeed, at $T=0$ (and fixed $t$), when $\mu-\mu^-_c$ goes from negative to positive, one induces a quantum phase transition between the vacuum of particles to a superfluid state (with finite density).\cite{SachdevBook}  Above two dimensions (the upper critical dimension of the $T=0$ quantum phase transition), the boson-boson (renormalized) interaction $\lambda_k$ is irrelevant and the critical behavior at the transition is mean-field, with a correlation-length exponent $\nu=1/2$ and a dynamical exponent $z=2$. However, $\lambda_k$ cannot be completely ignored and enters the equation of state (it is dangerously irrelevant  in the renormalization group sense). In fact, the thermodynamics close to the critical point $\mu=\mu^+_c=2dt$ can be interpreted in the same way, as the transition goes from the vacuum of holes (for $\mu>\mu_c^+$) to a superfluid with a finite density of holes (for $\mu<\mu_c^+$), the only difference being that the excitations are hole-like (with a negative quasi-particle weight). This has a direct correspondence with the density-driven Mott transition in the context of the Bose-Hubbard model. \cite{Fisher1989,Rancon2012,Rancon2012b} In particular, it implies that the thermodynamics is given by the universal functions of the so-called $d$-dimensional dilute gas universality class, and reads for the condensate and superfluid densities
\begin{equation}
\begin{split}
n_0(\mu,T)&=\left(\frac{\delta\mu^\pm}{4\pi t} \right)^{d/2} \calF_d\bigg(\frac{T}{\delta\mu^\pm},\tilde g(\delta\mu^\pm)\bigg)\,,\\
n_s(\mu,T)&=\left(\frac{\delta\mu^\pm}{4\pi t} \right)^{d/2} \calG_d\bigg(\frac{T}{\delta\mu^\pm},\tilde g(\delta\mu^\pm)\bigg)\,,
\end{split}
\end{equation}
where $\delta\mu^\pm=\mp(\mu-\mu_c^\pm)$ is positive in the superfluid phase. Here $\tilde g(\epsilon)$ is a renormalized interaction given by
\begin{equation}
\tilde g(\eps) = \llbrace 
\begin{array}{lcc}
8\pi \sqrt{a_3^2\eps/t} & \mbox{if} & d=3\, , \\
- \dfrac{4\pi}{\ln\left(\half\sqrt{a_2^2\eps/t}\right)+C} & \mbox{if} & d=2\, ,
\end{array}
\right.
\label{eq_gT}
\end{equation}
where $C$ is the Euler constant and $a_d$ is the $d$-dimensional s-wave scattering length. It can be computed for instance in the $U\to\infty$ limit of the Bose-Hubbard model ($U$ is the finite boson-boson on-site interaction) ~\footnote{See for example the Appendix E of Ref. \onlinecite{Rancon2011a}}
\begin{equation}
a_d = \llbrace 
\begin{array}{lcc}
\frac{1}{8\pi A} & \mbox{if} & d=3\, , \\[.1cm]
\frac{e^{-C}}{2\sqrt{2}} & \mbox{if} & d=2 \,,
\end{array}
\right.
\label{eq_scat}
\end{equation}
where $A=\frac{\sqrt{6}}{384 \pi^3}\Gamma\left(\frac{1}{24}\right)\Gamma\left(\frac{5}{24}\right)\Gamma\left(\frac{7}{24}\right)\Gamma\left(\frac{11}{24}\right)\simeq 0.1264$ (with $\Gamma(z)$ the Gamma function) is related to the third Watson's triple integral.  \cite{Borwein2013}

For $\tilde g(\delta\mu^\pm)\ll1$, \emph{i.e.} close enough to the quantum critical point, the universal functions $\calF_d$ and $\calG_d$ can be computed at $T=0$ using Bogoliubov theory and read ~\footnote{An extensive discussion of the dilute Bose gas universality class, its universal functions and their relationship with Bogoliubov theory is given in Ref. \onlinecite{Rancon2012} for $d=2$ and Ref. \onlinecite{Rancon2012b} for $d=3$.}
\begin{equation}
\begin{split}
\calF_3(0,y)=\frac{8\pi^{3/2}}{y}\Big(1-\frac{5\sqrt{2}y}{12\pi^2}\Big)\,,\\
\calG_3(0,y)=\frac{8\pi^{3/2}}{y}\Big(1-\frac{\sqrt{2}y}{3\pi^2}\Big)\,,
\end{split}
\end{equation}
in $d=3$, and 
\begin{equation}
\begin{split}
\calF_2(0,y)=\frac{4\pi}{y}+\frac{\ln2-2}{2}\,,\\
\calG_2(0,y)=\frac{4\pi}{y}+\frac{\ln2-1}{2}\,,
\end{split}
\label{eq_dil2D}
\end{equation}
in $d=2$. We have used the fact that the superfluid density is equal to the density of particles in Bogoliubov theory at zero temperature to obtain $\calG_d$. The dilute limit is also shown in Figures \ref{fig_T02D} and  \ref{fig_T03D}.\footnote{See also Ref. \onlinecite{Bernardet2002} for a discussion of the dilute limit in $d=2$.} We clearly see that it works only very close to the critical point $\delta\mu^\pm\ll t$, when the system is dilute enough, and one can neglect the effects of the lattice. We notice that, in two dimensions, the NPRG fails to reproduce correctly the dilute limit, unless the $\delta\mu^\pm$ is infinitesimally small (the vacuum limit $\delta\mu^\pm=0$ is correctly described). (However, Eq. \eqref{eq_dil2D} describes well the Monte Carlo data in the dilute regime.)  We ascribe this to a failure of the derivative expansion at this order in strong coupling, which could be corrected by including more terms in the expansion, or by working in a fully self-consistent BMW scheme.\footnote{The failure comes from $V_{A,k}$ which grows very fast in strong coupling for $d=2$, and perturbs the flow of $n_{0,k}$. This is not the case in $d=3$, where the flow of $V_{A,k}$ is much slower.}

\subsection{Finite temperature}

The NPRG allows us to study quantum systems at all temperature, and also successfully describes finite temperature phase transitions. Furthermore, it reproduces at least qualitatively the expected behavior of the BKT phase in two dimensions.

\subsubsection{Superfluid to normal fluid transition in $d=3$}
\begin{figure}[h!]
\centerline{\includegraphics[width=6cm]{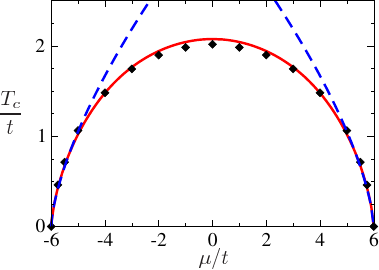}}
\caption{(Color online) Critical temperature as a function of the chemical potential in three dimensions. Symbols are Monte Carlo data from Ref. \onlinecite{Carrasquilla2012}. Dashed lines show the dilute limit, see Eq. \eqref{eq_Tcbogo}. } 
\label{fig_Tc} 
\end{figure}

To find the finite temperature phase diagram, we look for the lowest temperature such that $n_{0,k=0}=0$, which signals the transition between the Bose condensate and the normal phase. Figure \ref{fig_Tc} shows these results, which compare surprisingly well with the Monte Carlo calculation of Ref. \onlinecite{Carrasquilla2012}, the largest error being at most of the order of $3\%$ (close to $\mu=0$). The quantum critical points at $T=\delta\mu^\pm=0$ help one to understand the phase diagram away from the particle-hole symmetric point $\mu=0$. The critical temperature takes the scaling form
$$ T_c(\delta\mu^\pm)=\delta\mu^\pm \calH_d\big(\tilde g(\delta\mu^\pm)\big),$$
where to lowest order in $\tilde g(\delta\mu^\pm)$  in three dimensions ~\cite{Rancon2012b}
\begin{equation}
\calH_3(y)= \frac{4\pi}{\big(2\,\zeta(3/2)\,y\big)^{2/3}}\,,
\end{equation}
with $\zeta(x)$ is the Zeta function. This implies
\begin{equation}
\begin{split}
\frac{T_c}{t}&= 4\pi\bigg(\frac{A}{2\,\zeta(3/2)}\bigg)^{2/3}\bigg(\frac{\delta\mu^\pm}{t}\bigg)^{2/3}\,,\\
&\simeq 1.05 \bigg(\frac{\delta\mu^\pm}{t}\bigg)^{2/3}\,,
\label{eq_Tcbogo}
\end{split}
\end{equation}
in nice agreement with our calculations and the Monte Carlo for $\delta\mu^\pm\ll t$.
\begin{figure}[h!]
\centerline{\includegraphics[width=6cm]{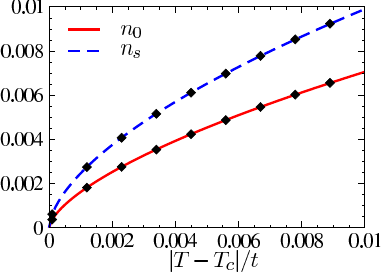}}
\caption{(Color online) Scaling of the condensate and superfluid densities close to the finite temperature phase transition at $\mu\simeq5t$ in three dimensions. Symbols: fits using Eq. \eqref{eq_scaling3D} using $\nu\simeq0.606$ and $\eta\simeq0.059$. } 
\label{fig_scaling3D} 
\end{figure} 

The NPRG also allows us to study the critical behavior close to the finite-temperature phase transition. It is well known that this transition is described by the three dimensional classical  $O(2)$  model, and that the RG flow goes to the (non-trivial) Wilson-Fisher fixed point. At the critical point, the condensate and  superfluid  densities scale as 
\begin{equation}
\begin{split}
n_{0,k}= \tilde n_0^* k^{d-2+\eta^*}\,,\\
n_{s,k}= \tilde n_s^* k^{d-2}\,,
\end{split}
\end{equation}
where $\eta^*\equiv\eta^*_A$ is the anomalous dimension at the fixed point ($\eta_{A,k}=-k\partial_k \ln Z_{A,k}$). Away from the fixed point in the ordered phase, the critical flow stops at a characteristic length scale, the Josephson length $\xi_J$,\cite{Josephson1966} scaling as $\xi_J\propto \alpha^{-\nu}$, where $\alpha\propto |T-T_c(\mu)|$ or $\alpha\propto |\mu-\mu_c(T)|$ depending on the control parameter, implying the scaling laws
\begin{equation}
\begin{split}
n_0\propto \alpha^{\nu(d-2+\eta^*)}\,,\\
n_s\propto \alpha^{\nu(d-2)}\,.
\label{eq_scaling3D}
\end{split}
\end{equation}
Figure \ref{fig_scaling3D} shows the scaling of $n_0$ and $n_s$ in the critical regime, which is well fitted by the critical exponent $\nu\simeq0.606$ and $\eta\simeq0.059$, that should be compared to the best estimates of the three dimensional $O(2)$ model: resummed pertubative calculations ($\nu\simeq 0.6700$, $\eta\simeq0.0334$),\cite{Pogorelov2008} Monte Carlo simulations ($\nu\simeq 0.6717$, $\eta\simeq0.0381$),\cite{Campostrini2006} NPRG in the BMW approximation ($\nu\simeq 0.674$, $\eta\simeq0.041$).\cite{Benitez2012} It has been shown that the value of the critical exponents improves when one increases the order of the truncation of the potential, and the order of the derivative expansion. \cite{Canet2003,Canet2003a}

\subsubsection{BKT transition \label{subsubsec_BKT} }

In this section we show how the BKT transition temperature $\Tkt$ can be estimated from the NPRG approach. For the classical $O(2)$ model, the NPRG reproduces most of the universal properties of the BKT transition.\cite{Grater1995,Gersdorff2001} In particular one finds a line of quasi-fixed points, which  enables us to identify a low-temperature phase ($T<\Tkt$), where the running of the superfluid density $ n_{s,k}=Z_{A,k}(n_{0,k})n_{0,k}$, after a transient regime, becomes very slow, implying a very large  correlation length $\xi$ (although not strictly infinite as expected in the low-temperature phase of the BKT transition). In this low-temperature phase, the anomalous dimension $\eta_{A,k}$ depends on the (slowly varying) 
superfluid density $n_{s,k}$, which takes its largest value $\sim 1/4$ when the RG flow crosses over to the disordered (long-distance) regime, and is then rapidly suppressed as $n_{s,k}$ further decreases. On the other hand, the essential scaling $\xi\sim e^{\const/(T-\Tkt)^{1/2}}$ of the correlation length above the BKT transition temperature $\Tkt$ is reproduced.\cite{Gersdorff2001} Thus, although the NPRG approach does not yield a low-temperature phase with an infinite correlation length, it nevertheless allows us to estimate the BKT transition temperature, and reasonable estimates have been obtained using the lattice NPRG in the two-dimensional classical XY model,\cite{Machado2010} the quantum $O(2)$ model,\cite{Rancon2013} the dilute Bose gas,\cite{Rancon2012a} and the Bose-Hubbard model.\cite{Rancon2013a}

\begin{figure}[th!]
\centerline{\includegraphics[width=7.3cm]{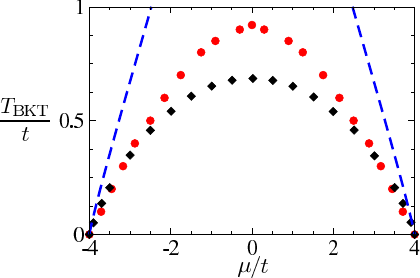}}
\caption{(Color online) Critical temperature $\Tkt$ as a function of the chemical potential in two dimensions (red circles). Black diamonds are Monte Carlo data from Ref. \onlinecite{Carrasquilla2012}. Dashed lines show the dilute limit Eq. \eqref{eq_Tktmc}.}
\label{fig_TKT}
\end{figure}

We use this method, described in detail in Ref. \onlinecite{Rancon2012a}, to compute the $\Tkt$ as a function of $\mu$, as shown in Fig. \ref{fig_TKT}. The results obtained  are of the correct order of magnitude compared to the Monte Carlo simulations of Ref. \onlinecite{Carrasquilla2012} and are up to $30\%$ off. This is not too surprising due to the crudeness of the present approximations, but is nevertheless encouraging for the method.  The dilute limit can be understood using the same approach as before. It has been shown that the BKT temperature of the two dimensional Bose gas is given by ~\cite{Popov_book_2,Fisher1988,Prokofev2001,Prokofev2002}
\begin{equation}
\left( \frac{\mu}{T}\right)_{\rm BKT}  = \frac{1}{2\pi}\tilde g\ln\left(\frac{2\zeta}{\tilde g}\right)\,, 
\label{eq_Tktmc}
\end{equation}
where $\zeta\simeq 13.2\pm 0.4$ has been obtained from a classical Monte Carlo simulation.\cite{Prokofev2001,Prokofev2002} This result, with the use of  Eqs. \eqref{eq_gT} and \eqref{eq_scat}, is also shown in Fig. \ref{fig_TKT}, and agrees with the Monte Carlo of Ref. \onlinecite{Carrasquilla2012} close to $\mu_c^\pm$. Here, as observed at zero temperature, the NPRG fails to reproduce the dilute limit.\footnote{It had been noted in Ref. \onlinecite{Rancon2012a} that the NPRG with the same approximations failed to reproduce the dilute limit for the BKT temperature in strong coupling for the Bose-Hubbard model. It does however work well in weak coupling.}

\section{Conclusion \label{sec_concl}}
We have presented a detailed Lattice NPRG study of the simplest model of quantum XY spins-$\half$  at zero and finite temperature using a mapping onto hardcore bosons. The Lattice NPRG allows us to take into account both the strong local correlations (imposed by the hardcore constraint) and the long-distance fluctuations, which are crucial in describing the critical behavior at finite temperature. The initial condition of the RG flow for hardcore bosons is described by the classical spin-wave theory for the corresponding XY model. The hardcore constraint imposes a very specific form for the functional behavior of the effective action and is conserved along the flow. We have therefore developed a field theoretic approach which is able to take into account strong constraints beyond a mean-field level. Since the Lattice NPRG is primarily  a renormalization group approach, it has the capability to describe non-trivial critical points, which in the present model exist only at finite temperature. Furthermore, the method reproduces most of the properties of the BKT phase in two dimensions. We have presented the simplest approximations and solved the corresponding flow equations. Already at this level, we obtained good results both at zero and finite temperature, for  both the thermodynamics and the critical behavior. Furthermore, the method can be improved by keeping more terms in the field expansion and the derivative expansion, which would increase the accuracy of the results for both the thermodynamics and the critical exponents.

 We have shown that the approach can reproduce the harmonic spin-wave corrections, which in the present model account for most of the quantum corrections for the thermodynamics, and compare very well with the Monte Carlo calculations.\cite{Coletta2012} However, the NPRG allows us to go beyond the $1/S$ expansion  and to describe quantum and classical critical regimes, which are out of reach of spin-wave theories. Although not discussed here, it has been shown that the NPRG is a method of choice in studying superfluid phases.\cite{Dupuis2007,Wetterich2008,Floerchinger2008,Dupuis2009,Dupuis2009a,Sinner2009,Sinner2010}
It is free of the infrared divergences usually encountered in perturbative approaches such as Bogoliubov or spin-wave theories, it satisfies the Hugenholtz-Pines theorem,\cite{Hugenholtz1959} and recovers the hydrodynamic regime at low energy (described by Popov's hydrodynamic theory,\cite{Popov1979} in the context of dilute superfluids). This is of particular importance for lattice systems away from the dilute limit.\cite{Rancon2011,Rancon2011a}

This work opens the door to the study of more complicated models presenting exotic phases. Indeed, the study of frustrated systems, for instance on the triangular or kagome lattices, and the predicted spin liquid phases would be an interesting challenge for the Lattice NPRG. This would certainly imply studying the momentum and frequency dependence of non-trivial correlation functions, which is possible in this formalism.\cite{Rancon2014} Furthermore, this approach allows for more refined initial conditions, including clusters of sites (instead of only one site as done here), as was already done in the fermionic language.\cite{Reuther2014,Kinza2013} It should be noted that our bosonic formulation is able to describe ordered phases, which are usually out of reach of the standard fermionic RG. \cite{Reuther2010,Reuther2011,Reuther2014}

Finally, we want to address the question of the treatment of the quantum Heisenberg model in this framework. This model is described, in the hardcore bosons formulation, by adding a nearest-neighbor interaction to the Hamiltonian \eqref{eq_H}. The standard, one-particle irreducible (1-PI), implementation of the NPRG, on which the current approach is based, assumes that the regulator term $\Delta \hat H_k$ is quadratic in the fields (see Eq. \eqref{eq_Hreg}), and thus does not allow for the decoupling of the nearest-neighbor interaction at the beginning of the flow. There exist however generalizations to the 2-PI case, which are important in describing ordered phases of fermionic systems,\cite{Dupuis2014} that would solve this problem. These formulations of the NPRG could be generalized to treat hardcore bosons, in  much the same spirit. (Note that this could also be of use for pseudo-fermion approaches.) In particular, the initial conditions would be the same (given by the on-site Hamiltonian solved exactly), the differences coming from the form of the regulator (both quartic and quadratic in the fields), as well as from the flow equations.

%\vspace{-0.5cm}
\begin{acknowledgments}
%\vspace{-0.3cm}
We thank J.  Carrasquilla for sharing the data of Ref. \onlinecite{Carrasquilla2012} and for interesting discussions, as well as N. Laflorencie for sharing the data of Ref. \onlinecite{Coletta2012}. Useful discussions with G. Uhrig are acknowledged. We particularly thank N. Dupuis and T. Roscilde for a critical reading of the manuscript. We are grateful to M. Valiente for pointing out that the three dimensional scattering length can be calculated analytically using Watson's triple integrals.
\end{acknowledgments}

%\newpage
%\vspace{-01.5cm}
\appendix

\section{Two-point vertex function and propagator \label{app_Gam}}

We discuss here some symmetries of the two-point vertex function $\Gamma^{(2)}$, and give the expression of the propagator $G=-\Gamma^{(2)-1}$. All these expressions are true for the scale-dependent effective action $\Gamma_k$, the (exact) effective action $\Gamma_{k=0}$, and the local effective action $\Gamma_\loc$ (and their corresponding effective potentials, two-point functions, and propagators), and we therefore suppress the index $k$ in the following expressions.

Due to the U(1) symmetry of the hardcore bosons Hamiltonian, the two-point vertex function in a constant field $\phi$ (see Eq. \eqref{eq_defG}), takes the form ~\cite{Dupuis2009a} 
\begin{equation}
\Gamma_{ij}^{(2)}(q;\phi) = \delta_{i,j}\,\Gamma_{A}(q;n) + \phi_i\phi_j\, \Gamma_{B}(q;n) + \eps_{ij}\, \Gamma_{C}(q;n) \, ,
\label{eq_Gam_app}
\end{equation}
where $n=|\phi|^2$, $q=(\q,i\w)$, the indices $i,j$ refer to the real and imaginary parts of $\phi=\frac{1}{\sqrt{2}}(\phi_1+i\phi_2)$, and $\eps_{ij}$ is the antisymmetric tensor. 

For $q=0$, we can relate $\Gamma^{(2)}_k$ to the derivative of the effective potential $V(n)$ (defined in Eq. \eqref{eq_effpot}), 
\begin{equation}
\Gamma^{(2)}_{ij}(q=0;\phi) = \frac{\partial^2 V(n)}{\partial\phi_i\partial\phi_j} = \delta_{i,j}V'(n) + \phi_i\phi_j V''(n)\,, 
\end{equation}
 ($V'(n)=\partial V/\partial n$, etc.)  so that
\begin{equation}
\begin{split}
\Gamma_{A}(q=0;n) &= V'(n)\, , \\
\Gamma_{B}(q=0;n) &= V''(n) \,, \\
\Gamma_{C}(q=0;n) &= 0\, . \\
\end{split}
\end{equation}
Furthermore, parity and time-reversal invariance imply\,\cite{Dupuis2009a}
\begin{equation}
\begin{split}
\Gamma_{A}(q;n) &= \Gamma_{A}(-q;n) = \Gamma_{A}(\q,-i\w;n)\, , \\ 
\Gamma_{B}(q;n) &= \Gamma_{B}(-q;n) = \Gamma_{B}(\q,-i\w;n)\, , \\ 
\Gamma_{C}(q;n) &= -\Gamma_{C}(-q;n) = - \Gamma_{C}(\q,-i\w;n)\, .
\end{split}
\label{sym}
\end{equation}

The propagator $G=-\Gamma^{(2)-1}$ can be expressed as 
\begin{align}
G_{ij}(q;\phi) ={}& \frac{\phi_i\phi_j}{2n} G_{\rm ll}(q;n) + \left( \delta_{i,j} -  \frac{\phi_i\phi_j}{2n} \right) G_{\rm tt}(q;n) \nonumber \\ & + \eps_{ij}  G_{\rm lt}(q;n) \,,
\end{align}
where ${\rm l}$ stands for longitudinal and ${\rm t}$ for transverse. We have
\begin{equation}
\begin{split}
G_{\rm ll}(q;n) &= - \frac{\Gamma_{A}(q;n)}{D(q;n)} \,, \\ 
G_{\rm tt}(q;n) &= - \frac{\Gamma_{A}(q;n)+2n\Gamma_{B}(q;n)}{D(q;n)}\, , \\ 
G_{\rm lt}(q;n) &= \frac{\Gamma_{C}(q;n)}{D(q;n)}\, ,
\end{split}
\label{green}
\end{equation}
with
$D=\Gamma_{A}^2+2n\Gamma_{A}\Gamma_{B}+\Gamma_{C}^2$. 

\vspace{4mm}

\section{Initial effective action at finite temperature\label{app_temp}}

\subsection{Effective potential}

At finite temperature, the local partition function and the order parameter in constant sources read
\begin{equation}
\begin{split}
Z_\loc(J,J^*)&=\left(2e^{\frac{\beta\mu}{2}}\cosh\frac{\beta\sqrt{4|J|^2+\mu^2}}{2}\right)^N\,,\\
\phi^{(*)}&= \frac{J^{(*)}}{\sqrt{4|J|^2+\mu^2}}\tanh\frac{\beta\sqrt{4|J|^2+\mu^2}}{2}\,.
\end{split}
\label{eq_ZphiT}
\end{equation}
The effective potential is obtained from 
\begin{equation}
V_\loc(n)=-\frac{1}{\beta N}\log Z_\loc(J,J^*)+J^*\phi+\phi^*J\, ,
\end{equation}
where the sources have to be expressed as functions of $\phi$ and $\phi^*$.
One has to inverse the relationship between the sources and the field $\phi$ numerically, and we therefore give the effective potential as an implicit function of $n=|\phi|^2$,
\begin{widetext}
\begin{equation}
\begin{split}
n&= \frac{|J|^2}{4|J|^2+\mu^2}\left[\tanh\frac{\beta\sqrt{4|J|^2+\mu^2}}{2}\right]^2\,,\\
V_\loc(n)&=-\frac{\mu}{2} -\frac{1}{\beta}\ln\left\{2\cosh\frac{\beta\sqrt{4|J|^2+\mu^2}}{2}\right\}+2\frac{|J|^2}{\sqrt{4|J|^2+\mu^2}}\tanh\frac{\beta\sqrt{4|J|^2+\mu^2}}{2}\,.
\label{eq_VnT}
\end{split}
\end{equation}
Figure \ref{fig_nVT}  shows $n$ as a function of the source and $V_\loc(n)$ for different temperatures.
\begin{figure}[h!]
\centerline{
\includegraphics[width=6cm]{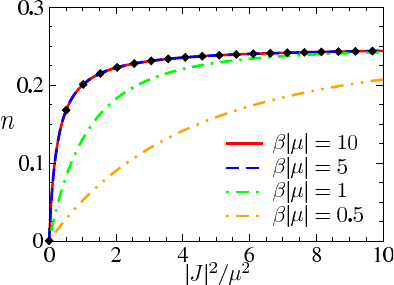}
\hspace{1cm}
\includegraphics[width=6.5cm]{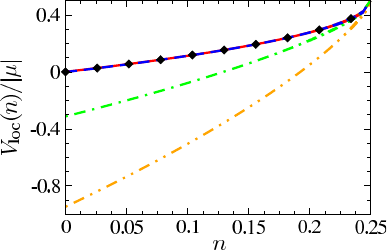}
}
\caption{(Color online) Left: $n=|\phi|$ as a function of the source $|J|^2$ for different temperatures.  Right: Local effective potential $V_\loc(n)$ as a function of $n$ for a negative chemical potential for different temperatures. Symbols show the zero temperature limit, see Eqs. \eqref{phiJ} and \eqref{eq_Vloc}. } 
\label{fig_nVT} 
\end{figure} 

In order to compute the coefficients of the field expansion of the effective potential $V_{0,\Lambda}$,  $\delta_\Lambda$ and $\lambda_\Lambda$ [see Eq. \eqref{trunc}], we also need the first two derivatives of the effective potential, which read
\begin{equation}
\begin{split}
V'_\loc(n)&=\sqrt{4|J|^2+\mu^2}\coth\frac{\beta\sqrt{4|J|^2+\mu^2}}{2}\,,\\
V''_\loc(n)&=\frac{2 \left(4 |J|^2+\mu ^2\right)^{3/2} \left(\sinh \left(\beta  \sqrt{4 |J|^2+\mu^2}\right)-\beta  \sqrt{4 |J|^2+\mu ^2}\right) \coth ^3\left(\frac{1}{2} \beta  \sqrt{4|J|^2+\mu ^2}\right)}{4 |J|^2 \beta  \sqrt{4 |J|^2+\mu ^2}+\mu ^2 \sinh \left(\beta  \sqrt{4|J|^2+\mu ^2}\right)}\,,
\label{eq_VpT}
\end{split}
\end{equation}
\end{widetext}
and which are  implicit functions of $n$ through Eq. \eqref{eq_VnT}. The initial effective potential at finite temperature is given by
\begin{equation}
V_\Lambda(n)=V_\loc(n)-2dt n\,.
\end{equation}
To obtain the initial value of the order parameter, we first need to find the value of the source $J_0$ such that $V'_\loc(n)-2dt=0$, that is
\begin{equation}
\sqrt{4|J_0|^2+\mu^2}\coth\frac{\beta\sqrt{4|J_0|^2+\mu^2}}{2}=2dt\,.
\label{eq_J0}
\end{equation}
Because the left-hand side of this equation is a monotonously  increasing function of $|J_0|^2$, the condition $V'_\loc(n)-2dt=0$ is impossible if
\begin{equation}
\mu\coth\frac{\beta\mu}{2}>2dt,
\end{equation}
implying $J_0=0$ and $n_{0,\Lambda}=0$, as well as
\begin{equation}
\begin{split}
V_{0,\Lambda}&=-\frac{\mu}{2} -\frac{1}{\beta}\ln\Big\{2\cosh\frac{\beta\mu}{2}\Big\}\,,\\
\delta_\Lambda&=\mu\coth\frac{\beta\mu}{2}-2dt\,,\\
\lambda_\Lambda&=2 \mu  \coth ^3\left(\frac{\beta  \mu }{2}\right) \left(1-\frac{\beta  \mu }{ \sinh\beta  \mu }\right)\,.
\end{split}
\end{equation}
The case 
$$\mu\coth\frac{\beta\mu}{2}=2dt$$
corresponds to the  mean-field phase transition between the superfluid and the normal phase,\cite{Matsubara1956} with $\delta_\Lambda=n_{0,\Lambda}=0$. This defines the mean-field critical temperature $T_c^{\rm MF}(\mu)$. Finally, if 
\begin{equation}
\mu\coth\frac{\beta\mu}{2}<2dt\,,
\end{equation}
one has to find $J_0$ numerically through Eq. \eqref{eq_J0} and compute $V_{0,\Lambda}$, $n_{0,\Lambda}$ and $\lambda_\Lambda$ using Eqs. \eqref{eq_VnT} and \eqref{eq_VpT} ($\delta_\Lambda=0$ by definition).

\subsection{Inverse propagator}

The calculation of the inverse propagator is mostly the same as in the case of zero temperature, but one must be careful in the handling of the zero-frequency terms. Indeed, at zero temperature, for the normal  propagator (and similarly for the anomalous propagator), the disconnected part $|\mean{\hat b}|^2$ is exactly compensated and no term proportional to $\beta \delta_{n,0}$ appears after Fourier transformation from imaginary time to Matsubara frequencies $\w_n=2\pi nT $. This is not the case at finite temperature. An explicit calculation of the Green functions at finite temperature as functions of the sources gives
\vspace{1cm}
\begin{widetext}
\begin{equation}
\begin{split}
G_{\rm n}(i\w_n)&=-\frac{\left(2 |J|^2+\mu  (\mu -i \w_n)\right) \tanh \left(\frac{1}{2} \beta  \sqrt{4 |J|^2+\mu
   ^2}\right)}{\sqrt{4 |J|^2+\mu ^2} \left(\w_n^2+4 J^2+\mu ^2\right)}-\beta \delta_{n,0}\frac{|J|^2   \text{sech}^2\left(\frac{1}{2} \beta  \sqrt{4 |J|^2+\mu ^2}\right)}{4 |J|^2+\mu^2}\,,\\
G_{\rm an}(i\w_n)&=\frac{2 J^2 \tanh \left(\frac{1}{2} \beta  \sqrt{4 |J|^2+\mu ^2}\right)}{\sqrt{4 |J|^2+\mu ^2}
   \left(\w_n^2+4 |J|^2+\mu ^2\right)} - \beta\delta_{n,0} \frac{J^2  \text{sech}^2\left(\frac{1}{2} \beta  \sqrt{4 |J|^2+\mu ^2}\right)}{4 |J|^2+\mu
   ^2}\,.
\end{split}
\label{eq_GT}
\end{equation}
Using Eq. \eqref{eq_GamG}, we obtain the three components of the inverse propagator
\begin{equation}
\begin{split}
\Gamma_{A,\loc}(i\w_n;n) &= \sqrt{4 |J|^2+\mu ^2} \coth \left(\frac{1}{2} \beta  \sqrt{4 |J|^2+\mu ^2}\right) \,, \\ 
\Gamma_{B,\loc}(i\w_n;n) &=\delta_{n,0} V''_\loc(n)+(1-\delta_{n,0})\frac{2 \left(4 |J|^2+\mu ^2\right)^{3/2} \coth ^3\left(\frac{1}{2} \beta  \sqrt{4 |J|^2+\mu
   ^2}\right)}{\mu ^2}\, , \\ 
\Gamma_{C, \loc}(i\w_n;n) &= -\frac{  \sqrt{4 |J|^2+\mu ^2} \coth \left(\frac{1}{2} \beta  \sqrt{4 |J|^2+\mu
   ^2}\right)}{\mu }\w_n\,.
\end{split}
\end{equation}
\end{widetext}
While at finite temperature $\Gamma_{A,\loc}$ stays frequency independent (and equal to $V'_\loc(n)$) and $\Gamma_{C,\loc}$ is still linear in frequency (defining a finite temperature $Z_{C,\loc}$), we see that $\Gamma_{B,\loc}$ now has a very peculiar frequency dependence: it is equal to $V''_\loc(n)$ for $\w_n=0$, as it should, but takes another (constant) value for all $\w_n\neq 0$. This is nevertheless consistent with the high frequency behavior of the propagators, see Appendix \ref{app_highw}. Of course, in the limit $\beta\to\infty$, we recover the zero temperature limit $\Gamma_{B,\loc}(i\w;n)=V''_\loc(n)$ for all frequencies.

\section{High frequency behavior of the local Green functions\label{app_highw}}

\subsection{Zero temperature}
The normal and anomalous local Green functions defined in Eq. \eqref{eq_Green} can be written as
\begin{equation}
\begin{split}
G_{\rm n}(i\w)&=\int_{-\infty}^\infty d\w' \frac{A_{\rm n}(\w')}{i\w-\w'}\,,\\
G_{\rm an}(i\w)&=\int_{-\infty}^\infty d\w' \frac{A_{\rm an}(\w')}{i\w-\w'}\,,
\end{split}
\label{eq_spec}
\end{equation}
when $i\w\neq 0$. The local spectral functions are defined by
\begin{equation}
\begin{split}
A_{\rm n}(t)&= \frac{1}{2\pi}\mean{\big[\hat b(t),\hat b^\dag(0)\big]}\,,\\
A_{\rm an}(t)&=\frac{1}{2\pi}\mean{\big[\hat b(t),\hat b(0)\big]}\,,
\end{split}
\end{equation}
where $\hat b(t)$ and $\hat b^\dag(t)$ are the hardcore boson operator in the Heisenberg picture, with local Hamiltonian $\hat H_\loc(J,J^*)$ (Eq. \eqref{eq_Hloc}). Form the spectral representation Eq. \eqref{eq_spec}, we obtain the high frequency expansion
\begin{equation}
\begin{split}
G_{\rm n}(i\w)&=\frac{a}{i\w}+\frac{b}{(i\w)^2}+O(\w^{-3})\,,\\
G_{\rm an}(i\w)&=\frac{c}{(i\w)^2}+O(\w^{-3})\,,
\end{split}
\end{equation}
where
\begin{equation}
\begin{split}
a&=\int_{-\infty}^\infty d\w A_{\rm n}(\w)\,,\\
 &=\mean{[\hat b,\hat b^\dag]}\,,\\
b&=\int_{-\infty}^\infty d\w \,\w A_{\rm n}(\w)\,,\\
 &=2\pi \Big[i\partial_t A_{\rm n}(t)\Big]_{t=0}=\mean{\big[[\hat b,\hat H_\loc],\hat b^\dag\big]}\,, \\
 c&=\int_{-\infty}^\infty d\w \,\w A_{\rm an}(\w)\,,\\
 &=2\pi \Big[i\partial_t A_{\rm an}(t)\Big]_{t=0}=\mean{\big[[\hat b,\hat H_\loc],\hat b\big]}\, .
\end{split}
\label{eq_spec2}
\end{equation}
To obtain Eq. \eqref{eq_spec2}, we used the equations of motion of the operators $\hat b(t)$ and $\hat b^\dag(t)$. Using the hardcore bosonic commutation relations, we obtain
\begin{equation}
\begin{split}
a&=1-2\mean{\hat b^\dag \hat b}\,,\\
b&=-\mu (1-2\mean{\hat b^\dag \hat b})+2 J \mean{\hat b^\dag}\,,\\
 c&=-2 J\mean{\hat b}.
\end{split}
\label{eq_spec3}
\end{equation}
The (field dependent) local density  $\bar n_\loc=\mean{\hat b^\dag \hat b}$ is given by 
\begin{equation}
\begin{split}
\bar n_\loc &=-\partial_\mu V_\loc(n)\,,\\
&= \half\big(1+{\rm sgn}(\mu)\sqrt{1-4n}\big)\,.
\end{split}
\end{equation}
Using the relationship between the field $\phi^{(*)}=\mean{\hat b^{(\dag)}}$ and the source $J^{(*)}$ 
\begin{equation}
\begin{split}
 \phi^{(*)}&=\frac{J^{(*)}}{\sqrt{4|J|^2+\mu^2}}\,,\\
 |J|^2&= \mu^2 \frac{n}{1-4n}\,,
\end{split}
\end{equation}
we obtain
\begin{equation}
\begin{split}
a&=-{\rm sgn}(\mu)\sqrt{1-4n}\,,\\
 &=\frac{1}{Z_{C,\loc}(n)}\,,\\
b&=|\mu|\left(\sqrt{1-4n}+\frac{2n}{\sqrt{1-4n}}\right)\\
 &=\frac{V'_\loc(n)+n V''_\loc(n)}{Z_{C,\loc}(n)^2}\,, \\
 c&=-\frac{2|\mu| \phi^2}{\sqrt{1-4n}}\,,\\
 &= -\frac{ \phi^2 V_\loc''(n)}{Z_{C,\loc}(n)^2}\,.
\end{split}
\end{equation}
These results are consistent with the high frequency behavior of the normal and anomalous local Green functions, see Sec. \ref{subsec_initial}, which are given by
\begin{equation}
\begin{split}
G_{\rm n}(i\w)&=-\frac{\Gamma_{A,\loc}(i\w;n)+n\Gamma_{B,\loc}(i\w;n)+i\Gamma_{C,\loc}(i\w;n)}{D_\loc}\,,\\
G_{\rm an}(i\w)&=\frac{\phi^2\Gamma_{B,\loc}(i\w;n) }{D_\loc}\,,
\end{split}
\label{eq_GGam}
\end{equation}
where $$D_\loc= \Gamma_{C,\loc}^2+\Gamma_{A,\loc}\big(\Gamma_{A,\loc}+2n\Gamma_{B,\loc}\big)\,,$$
the high frequency limit of which is given by 
$$D_\loc= \big(Z_{C,\loc}(n) \w\big)^2\,.$$

\subsection{Finite temperature}

At finite temperature, Eqs. \eqref{eq_spec},  \eqref{eq_spec2} and  \eqref{eq_spec3} are still valid. With the help of Eq. \eqref{eq_ZphiT}, one can compute $a$, $b$ and $c$ using
\begin{equation}
\begin{split}
\bar n_\loc&=\frac{1}{\beta} \frac{\partial \ln Z_\loc}{\partial \mu}\,,\\
&=\half+\frac{\mu  \tanh \left(\frac{1}{2} \beta  \sqrt{4 |J|^2+\mu ^2}\right)}{2\sqrt{4 |J|^2+\mu^2}}\,,
\end{split}
\end{equation}
as well as the results of appendix \ref{app_temp}. The results of an explicit calculation can be rewritten as
\begin{equation}
\begin{split}
a&=\frac{1}{Z_{C,\loc}(n)},\\
b&=\frac{V'_\loc(n)+n \Gamma_{B,\loc}(i\w_n;n)}{Z_{C,\loc}(n)^2}\,, \\
 c&= -\frac{ \phi^2 \Gamma_{B,\loc}(i\w_n;n)}{Z_{C,\loc}(n)^2}\,.
\end{split}
\end{equation}
Here $\Gamma_{B,\loc}(i\w_n;n)$ is evaluated at (large) finite frequencies, and is therefore not equal to $V''_\loc(n)$ (see Appendix \ref{app_temp}). These results are consistent with high frequency dependence of the Green functions (see Eq. \eqref{eq_GGam}).

\vspace{4mm}
\section{One-loop corrections at $T=0$\label{app_1loop}}
\subsection{Effective potential and condensate density}
In the standard implementation NPRG, one obtains the one-loop corrections by solving the flow equation \eqref{wetteq} without renormalizing the effective action in the right-hand side of the equation, which can be rewritten as
\begin{equation}
\partial_k \Gamma_k=\half \partial_k \Tr \log\Big\{\Gamma_\Lambda^{(2)}+R_k \Big\}\,.
\end{equation}
 This equation can be integrated exactly and gives 
\begin{equation}
 \Gamma_{1l}= \Gamma_\Lambda+\half\Tr \log\bigg\{ \frac{\Gamma_\Lambda^{(2)}}{\Gamma_\Lambda^{(2)}+R_\Lambda}\bigg\}\,.
 \label{eq_Gam1l}
 \end{equation}
Note the presence of the term $\Gamma_\Lambda^{(2)}+R_\Lambda=\Gamma^{(2)}_{\rm loc}$, which is important  in order to get the correct results. From this, we get the one-loop effective potential by evaluating the effective action in constant field and performing the integral over frequency, \cite*{[{See for example the Appendix G of }] Diener2008}
\begin{widetext} 
\begin{equation}
 V_{1l}(n)= V_\Lambda(n)+\frac{1}{2 |Z_{{\rm loc},C}|}\sum_\q\bigg\{ \sqrt{\Big(t_\q+V'_{\rm loc}+nV''_{\rm loc}\Big)^2- \Big(n V''_{\rm loc}\Big)^2}-\sqrt{\Big(V'_{\rm loc}+nV''_{\rm loc}\Big)^2- \Big(nV''_{\rm loc}\Big)^2}-t_\q\bigg\}\,,
 \end{equation}
\end{widetext} 
where we have hidden the field dependence of $Z_{{\rm loc},C}$ and $V_{\rm loc}$ for notational convenience.

To obtain the one-loop pressure and condensate density, we rewrite the effective potential as $V_{1l}=V_\Lambda+\delta V$ and the condenstate density as $n_{0,1l}=n_{0,\Lambda}+\delta n_0$. Then $n_{0,1l}$ is defined by $V'_{1l}(n_{0,1l})=0$, which gives
\begin{equation}
\begin{split}
V'_{1l}(n_{0,1l})=& V'_\Lambda(n_{0,1l})+\delta V'(n_{0,1l})\,,\\
=& V'_\Lambda(n_{0,\Lambda})+\delta n_0 V''_\Lambda(n_{0,\Lambda})+\delta V'(n_{0,\Lambda})+\cdots\,,\\
=& \delta n_0 V''_\Lambda(n_{0,\Lambda})+\delta V'(n_{0,\Lambda})+\cdots\,,\\
=&0,
\end{split}
\end{equation}
where the dots stand for higher loop corrections. From this, we get
\begin{equation}
\begin{split}
n_{0,1l}&=n_{0,\Lambda}-\frac{\delta V'(n_{0,\Lambda})}{V''_\Lambda(n_{0,\Lambda})}\,,\\
		&=n_{0,\Lambda}-\sum_\q\bigg\{ \frac{(2dt)^2+dt(1-\bar\mu^2) t_\q-\bar\mu^2 t_\q^2}{4dt E_\q}-\half \bigg\}\,,
\end{split}
\end{equation}
where we have used Eq. \eqref{eq_dispMF} and $\bar\mu=\mu/(2dt)$.
This is the result quoted by Coletta \emph{et al.} in Ref \onlinecite{Coletta2012}. (It is different from that of Ref. \onlinecite{Bernardet2002}, as discussed in Ref. \onlinecite{Coletta2012}.)  Note that the field dependence of the effective potential $V_\loc(n)$ and of $Z_{C,\loc}(n)$ is crucial in recovering the correct one-loop result. If we approximate $Z_{C,\loc}(n)$ by a constant and expand  $V_\loc(n)$ to order $n^2$, the error is of about $5\%$ at $\mu=0$.

The pressure at one-loop can be obtain from \
\begin{equation}
\begin{split}
P_{1l}&=-V_{1l}(n_{0,1l})\,,\\
		&=-V_{1l}(n_{0,\Lambda})-\delta n_0 V'_{1l}(n_{0,\Lambda})+\cdots\,,\\
		&=-V_{1l}(n_{0,\Lambda})-\delta n_0 V'_\Lambda(n_{0,\Lambda})+\cdots\,,\\
		&=-V_{1l}(n_{0,\Lambda}),
\end{split}
\end{equation}
which gives
\begin{equation}
P_{1l}=P_\Lambda-\half\sum_\q\big(E_\q-2dt\big)\,,
\end{equation}
  where we have used $\int_\q t_\q=0$. In the case $d=2$, this is the $1/S$ spin-wave correction given in Ref. \onlinecite{Coletta2012}. We therefore recover the correct result for the density, given by a derivative with respect to the chemical potential.

\subsection{Superfluid density}

The superfluid density is given by $n_s=Z_A n_0$, where $Z_A$ is obtained from 
\beq
Z_A=\frac{1}{t}\partial_{p_x^2}\Gamma^{(2)}_{22}(p;n_0)\Big|_{p=0},
\eeq
where $i=2$ is the transverse component (with the choice $\phi_i=\sqrt{2n_0} \,\delta_{i,1}$).
We now show that we can recover the one-loop superfluid density at the beginning of the flow. First, let us remark that at one loop 
\beq
n_{s,1l}=Z_{A,1l}\,n_{0,1l}=n_{0,1l}+\delta Z_A n_{0,\L}
\eeq
where $Z_{A,1l}=Z_{A,\L}+\delta Z_A$ and we have used $Z_{A,\L}=1$. Since we have already computed $n_{0,1l}$ above we only need to compute $\delta Z_A$.

The one-loop correction of the inverse propagator is obtained from (see Eq. \eqref{eq_Gam1l})
\beq
\delta \Gamma=\half \Tr\log \left\{\frac{\Gamma^{(2)}_\Lambda}{\Gamma^{(2)}_\Lambda+R_\Lambda}\right\}.
\eeq
Since $\Gamma^{(2)}_\Lambda(q)+R_\Lambda(\q)=\Gamma^{(2)}_{\rm loc}(\w)$ is independent of momenta, and since the only momentum dependence of $\Gamma_\Lambda$ comes from $\Delta H_\L[\phi^*,\phi]$, all the $n$-point vertices with $n>2$ are momentum independent
\beq
\Gamma^{(n>2)}_{i_1...i_n,\Lambda}(\{q_i\};n)=\Gamma^{(n>2)}_{i_1...i_n,\loc}(\{\w_i\};n),
\eeq
which will greatly simplify the calculation. In particular, the term $\half \Tr\log \Gamma^{(2)}_{\rm loc}$ is purely local and
will not contribute to $\delta Z_A$, and we will therefore not calculate its contribution to $\delta\Gamma^{(2)}$. Furthermore, when computing $\delta \Gamma^{(2)}$, there are two pieces that can in principle contribute to its momentum dependence. The first one (here and below $p=(0,\p)$ and $q=(\w,\q)$)
\begin{widetext} 
\beq
\begin{split}
\Big(\delta\Gamma_{ij}(p;n)\Big)_1&=\half\sum_{i_1,i_2}\int_q\left\{ \Gamma_{i j i_1 i_2,\L}^{(4)}(p,-p,q,-q;n) \,G_{i_1 i_2,\L}(q;n)\right\},\\
&=\half \sum_{i_1,i_2}\int_q\left\{ \Gamma_{ij i_1 i_2,{\rm loc}}^{(4)}(0,0,\w,-\w;n) \,G_{i_1 i_2,\L}(q;n) \right\},
\end{split}
\eeq
is momentum independent and can be ignored. 
 The second contribution is 
\beq
\begin{split}
\Big(\delta\Gamma_{ij}(p;n)\Big)_2&=-\half \sum_{i_1,i_2,i_3,i_4}\int_q\left\{ \Gamma_{i i_1 i_2,\L}^{(3)}(p,q,-p-q;) \,G_{i_1 i_3,\L}(q;n) \,G_{i_2 i_4,\L}(-p-q;n)\,\Gamma_{i_3 i_4 j,\L}^{(3)}(-q,p+q,-p;n)\right\},\\
&=-\half \sum_{i_1, i_2, i_3, i_4}\int_q\left\{ \Gamma_{i i_1 i_2,{\rm loc}}^{(3)}(0,\w,-\w;n)\, G_{i_1 i_3,\L}(q;n) \,G_{i_2 i_4,\L}(-p-q;n)\,\Gamma_{i_3 i_4j,{\rm loc}}^{(3)}(-\w,\w,0;n)\right\}.
\end{split}
\eeq
%\end{widetext} 
Using $\Gamma^{(3)}_{ijl}(0,q,-q;n)=(\beta N)^{-\half}\partial_{\phi_i}\Gamma^{(2)}_{jl}(q;n)$, \cite{Blaizot2006} and Eq. \eqref{eq_Gam_app}, we find
\begin{eqsplit}
\Gamma_{i i_1 i_2,{\rm loc}}^{(3)}(0,\w,-\w;n)=\phi_i\left(\delta_{i_1,i_2}\Gamma_{A,\loc}'(\w;n)+\phi_{i_1}\phi_{i_2}\Gamma_{B,\loc}'(\w;n)+\epsilon_{i_1 i_2}\Gamma_{C,\loc}'(\w;n)\right)+(\delta_{i, i_1}\phi_{i_2}+\delta_{i ,i_2}\phi_{i_1})\Gamma_{B,\loc}(\w;n),
\end{eqsplit}
where $\Gamma_{\alpha,\loc}'(\w;n)=\partial_n\Gamma_{\alpha,\loc}(\w;n)$. Using Eqs.  \eqref{eq_Vloc} and \eqref{eq_G2loc}, as well as $\phi_2=0$, we obtain 
\beq
\Gamma_{2ij,{\rm loc}}^{(3)}(0,\w,-\w;n_{0,\L})=\frac{\lambda_\L\sqrt{2 n_{0,\L}}}{\sqrt{\beta N}}\left(\delta_{i,2}\delta_{j,1}+\delta_{j,2}\delta_{i,1}\right),
\eeq
with
\beq
\begin{split}
\lambda_\L&=V''_\L(n_{0,\L}),\\
&=\frac{4dt}{\bar\mu^2},
\end{split}
\eeq
which is frequency independent.
We thus get
\beq
\delta Z_A = -\frac{2\lambda_\L^2 n_{0,\L}}{t}\partial_{p_x^2}  \int_q\Big\{ \bar G_{ll}\left(p+q\right)\bar G_{tt}\left(q\right)+\bar G_{lt}\left(p+q\right)\bar G_{lt}\left(q\right)   \Big\}\bigg|_{\p={\bf 0}},
\eeq
\end{widetext}
where $\bar G_{ij}(q)=G_{i_1 i_3,\L}(q;n_{0,\L})$, see Eq. \eqref{green}. Using
\begin{eqsplit}
\partial_{p_x^2}\int_q \bar G_\alpha(q) \bar G_\beta(p+q)\Big|_{p=0}&=\half\int_q \bar G_\alpha(q) \partial^2_{p_x}\bar G_\beta(p+q)\Big|_{p=0},\\
&= -\half\int_q \partial_{q_x}\bar G_\alpha(q) \,\partial_{q_x}\bar G_\beta(q),\\
&=-\half\int_q \left(\frac{\partial\eps_\q}{\partial q_x}\right)^2 \bar G'_\alpha(q)\,\bar G'_\beta(q),
\end{eqsplit}
where we have used an integration by part to go from the first line to the second, and $\bar G'_\alpha(q)=\partial_{\eps_\q}\bar G_\alpha(q)$. We thus obtain
\begin{eqnarray}
\delta Z_A &= \frac{\lambda_\L^2 n_{0,\L}}{t}\int_q \left(\frac{\partial\eps_\q}{\partial q_x}\right)^2 \left[\bar G'_{11}(q)\bar G'_{22}(q)+\bar G'_{12}(q)\bar G'_{12}(q)\right],\nonumber\\
&=\frac{\lambda_\L^2 n_{0,\L}}{4t|\bar Z_C|}\int_\q \frac{\left(\frac{\partial\eps_\q}{\partial q_x}\right)^2}{\left(\eps_\q(\eps_\q+2\lamb_\L n_{0,\L}\right)^{3/2}},
\end{eqnarray}
where we have performed the integral over frequency on the second line and $\bar Z_C = Z_{C,\L}(n_{0,\L})=-1/\bar\mu$. Putting everything together, we obtain for the hypercubic lattice
\beq
n_{s,1l}=n_{0,1l}+d^2(1-\bar\mu^2)^2 t^3\int_\q \frac{\sin^2(q_x)}{E_\q^3}.
\eeq
One can check, at least numerically, that this result is in perfect agreement with that of Coletta \emph{et al.} \cite{Coletta2012} in two dimensions.

\bibliography{../../../Articles/bibli_bosons,../../../Articles/bibli_RG,../../../Articles/bibli_disorder,../../../Articles/bibli_OutEq,../../../Articles/bibli_BCSBEC,../../../Articles/bibli_polar,../../../bibli_diverse}

\end{document}